\documentclass[aps,pre,twocolumn,superscriptaddress,10pt,%
floatfix]{revtex4-1}
\usepackage{graphicx}
\usepackage{amsmath}
\usepackage{natbib}
\usepackage[utf8]{inputenc}
\usepackage{amssymb,nicefrac}
\usepackage{color}
\usepackage{newfloat,microtype}

\definecolor{cream}{RGB}{222,217,201}
\usepackage{microtype}

\definecolor{cream}{RGB}{222,217,201}
\usepackage{microtype}

\begin{document}

\title{
    Snapping elastic disks as  microswimmers:
    swimming at low Reynolds numbers
     by shape hysteresis  }

 \author{Christian Wischnewski}
 \affiliation{Physics Department, TU Dortmund University, 44221 Dortmund,
   Germany} 
 \author{Jan Kierfeld}
\affiliation{Physics Department, TU Dortmund University, 44221 Dortmund,
  Germany}

\email{jan.kierfeld@tu-dortmund.de}

\begin{abstract}
We illustrate a concept for shape-changing microswimmers, which
  exploits the hysteresis of a shape transition of an elastic object,
   by an elastic disk undergoing cyclic localized swelling.
Driving the control parameter of  a hysteretic shape transition
in a completely time-reversible manner gives rise to a non-time-reversible
     shape sequence and a net swimming
motion if the elastic object is immersed into a viscous fluid.
We prove this concept with a microswimmer
which is a flat circular elastic disk that  undergoes a 
transition into a dome-like   shape by  localized swelling
of an inner disk.
The control parameter of this shape transition is a scalar
swelling factor of the disk material.
With  a fixed outer frame with an additional
attractive interaction in the central region, the
shape transition between
     flat  and  dome-like  shape becomes hysteretic and resembles a
    hysteretic  opening and closing of a scallop.
Employing Stokesian dynamics simulations of a discretized
version of the disk we show
 that the swimmer is effectively moving into
 the direction of the opening of the dome in a viscous fluid
if the swelling parameter is changed in a time-reversible manner.
The swimming mechanism can be  qualitatively reproduced by a
simple  9-bead model.                                      
\end{abstract}

\maketitle

\section{Introduction}

Swimming on the microscale at low Reynolds numbers requires special propulsion
mechanisms which are effective in the presence of dominating viscous
forces.
Phoretic swimmers  create gradients in external fields such as
concentration or temperature which in turn give rise
to symmetry-breaking interfacial forces leading to propulsion
if they overcome the friction force of the microswimmer
\cite{Illien2017}.
Besides phoretic swimmers, self-deforming or shape-changing
swimmers are the largest class
of microswimmers. They deform their body in a cyclic way in order to propel.
This general principle has the advantage that it works 
independently of the environment, i.e., it does not require an external
field, that eventually changes the properties of the fluid. 
The disadvantage of swimming by deformation is, on the other
hand,  that there are necessarily ``moving parts'' causing additional
viscous flow  in the fluid and forces on the swimmer.
As a consequence, at low Reynolds numbers, 
the cyclic deformation pattern must not be invariant under time-reversal:
the scallop theorem formulated by Purcell states that
periodic reciprocal  patterns of deformation can not lead to an effective
net motion  on the microscale because of
the linearity of the Stokes equation \cite{Purcell1977}.

In nature, many
different examples of deformation swimmers can be found such as  bacteria,
algae and spermatozoa \cite{Lauga2009,Elgeti2015}. These natural swimmers often
rely on the movement of a few flagella or many cilia on their surface
\cite{Taylor1951,Berg1973,Goldstein2015,Jeanneret2016}.
Flagella  employ a periodic forcing but overcome the scallop theorem by
exploiting friction along the elastic flagellum to break 
time-reversibility.
This requires a matching of 
driving and frictional damping time scales for 
efficient propulsion. Often it also requires 
the ability of local actuation for the periodic forcing.
This makes this concept hard to reproduce or imitate in a controlled
fashion in an 
artificial system \cite{Dreyfus2005,Tottori2013}.

Another  basic strategy to overcome the scallop theorem are
 deformation cycles that involve at least two control parameters
and drive the swimmer periodically along a closed contour
in this at least  two-dimensional parameter space.
Different shape changing 
artificial swimmers have been
developed based on this concept starting with 
 Purcell's three-link swimmer \cite{Purcell1977} and including
 swimmers performing
 small deformations of spheres, circles, or cylinders
\cite{Lighthill1952,Blake1971,shapere1987,Felderhof1994,avron2004},
or   shape-changing 
vesicles \cite{evans2010}.
The most simple  shape changing microswimmer is arguably the
one-dimensional linear three-bead swimmer developed by Najafi and Golestanian,
where three beads change their distance in a  non-time-reversible
way \cite{NajafiGolestanian2004,Golestanian2008}. By extending the linear
three bead arrangement to a second dimension in
a triangular shape,  a three bead swimmer can perform
two-dimensional motions (circles) \cite{Ledesma-Aguilar2012} and 
steer \cite{Rizvi2018}.
Nevertheless, despite the simplicity of the
concept,  this type of  swimmer is difficult to implement
experimentally because it  requires fine control 
over, at least, two control parameters such as the bead positions 
of the three-bead swimmer \cite{Leoni2009,Golestanian2010}.

\begin{figure}
	\centering
  \includegraphics[width=0.9\linewidth]{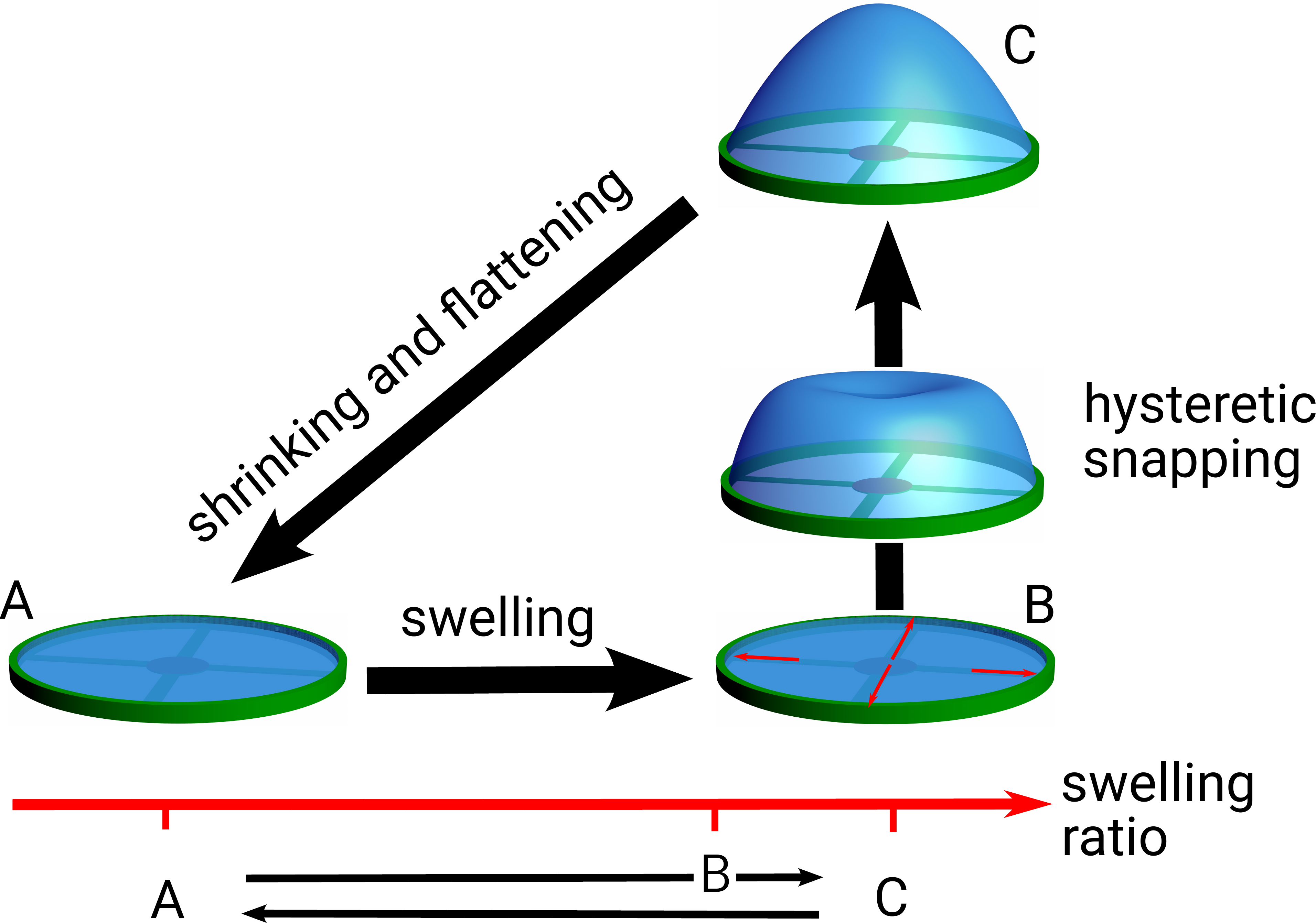}
    \caption{A completely time-reversible oscillation of the control parameter
      (horizontal axis, here: swelling ratio)
      gives rise to  non-time-reversible
      shape cycle of an elastic disk
       because of hysteresis of the triggered shape
       transition. The deformation cycle between shapes A and C
       resembles the
       opening and closing of a scallop but is hysteretic (compressed
        shape B and the transition state between B and C are only
        visited upon swelling).
      }
		\label{fig:hysteresis_sketch}
\end{figure}

We employ a different general strategy in order to
overcome the scallop theorem, which is 
widely applicable and only involves  control of a single
global and scalar
control parameter, which couples, however,  to a  hysteretic
(or  bistable)
shape transition of the system, see Fig.\ \ref{fig:hysteresis_sketch}.  
If also the sequence of shapes exhibits hysteresis, this
converts the  time-reversible 
motion in   one-dimensional control parameter space into 
a non-time-reversible motion in a higher-dimensional 
parameter or shape space. 
Hysteretic shape transitions can be realized, for example,
by using the  intrinsic properties of elastic
materials. In this work, we will realize
such a hysteretic shape transition 
based on a swelling process of a flat and thin circular elastic disk, where
material swelling with swelling ratios of only a few percent
in the central region of the disk
leads to  a shape transition from the flat disk shape  into  curved
conformations, such as a dome-like shape
\cite{KleinEfrati2007,Efrati2009,Pezulla2015}. The snapping into
an elliptic dome-like shape actually faintly resembles  the opening
and closing of a
scallop. By  further enhancing the elastic disk 
with   a fixed frame with attractive interactions
we can endow this transition with genuine hysteretic effects. 
These hysteresis effects  allow us to break the reciprocity
of the shape cycle 
although we employ simple cyclic and   fully time-reversible
oscillations of 
the swelling factor as    single global and scalar
control parameter.
The main point of this paper is to give  the  proof of concept
that this leads to net propulsion and is a viable
realization of a microswimmer. 
The principle of exploiting a periodically driven
  hysteretic shape transition
  in order to achieve net propulsion has been introduced
  in Ref.\  \citenum{Djelloul2017} using the 
the buckling transition of spherical elastic
shells as propulsion mechanism.
The buckling of spherical shells 
is  a subcritical hysteretic shape transition
\cite{Knoche2011a,Knoche2014,Baumgarten2019},
which will turn out to be conceptually similar to the
elastic instability triggered in the elastic disk by localized
swelling.

Deformation cycles of 
elastic materials have been applied before to design
artificial swimmers.
In Ref.\  \citenum{Palagi2016}, structured light fields were
used to drive elastic deformation waves on swimmers with a
homogeneous body made of a soft material.
Other approaches focused on  elastic
double layers, where swelling of one layer can induce bending;
such externally controllable
swelling layers can be engineered, for example, using
thermoresponsive microgels \cite{Mourran2017}.
These ideas 
were used to design swimmers with a
helical structure that can be propelled by conformation changes of the
helix \cite{Mourran2017,Zhang2017,Koens2018}. Here conformation changes
are  non-reciprocal, partly because of hysteresis effects in  the
heating cycle of the thermoresponsive gel.

Deformation swimmers are relatively slow in general, because the swimming
distance only scales quadratically with  the deformation
displacement for many
deformation swimmers \cite{Lighthill1952,Blake1971}.
Therefore, a high frequency of  conformation
changes is needed in oder to achieve a significant swimming velocity.
This applies also to the concept of swimming by hysteretic shape changes.  
The driving frequency is, however,  not limited by an  
additional damping time scale (as, for example, in flagellar motion)
because breaking 
of the time-reversibility is inherent 
in the hysteretic shape sequence itself but,
at high driving frequencies, one could leave the realm of
low Reynolds numbers. 

The paper is organized as follows. 
At first, we  present a ``dry''
numerical analysis of the elastic deformation cycle of the elastic disk
with material swelling in the interior in the absence of a surrounding fluid and
corresponding hydrodynamic interactions.
We quantify the hysteretic effects of the swelling transition and
show that there is most likely  no genuine  hysteresis in
experiments on simple swelling disks.
We can enhance the model system by adding an additional fixed frame with an
attractive interaction to the disk in order to generate a robust and
genuine hysteretic shape transition.
Then we perform  a ``wet'' hydrodynamic simulation featuring a Rotne-Prager
interaction in order to proof the swimming ability of this system and
characterize the net propulsion. Finally, a
simplified 9-bead model is presented that mimics  the essence of the 
underlying swimming mechanism and is able to qualitatively reproduce and
explain its main characteristics.

\section{Swelling of a flat disk}

\subsection{Theory}

In the following, we consider a flat circular disk with radius
$R_\mathrm{out}$ and  thickness $h$. The disk shall be very thin
($h\ll R_\mathrm{out}$), so we can use a  two-dimensional model. We
parametrize our two-dimensional disk in polar coordinates $(r,\,\varphi)$. The
basic idea is to deform the disk into a curved shape
by a localized, i.e., inhomogeneous  swelling process 
which changes the disk's metric
\cite{KleinEfrati2007,Efrati2009,Pezulla2015}.
By swelling we mean  a local isotropic swelling, where 
 the rest lengths of fibers change by a position-dependent factor
 $A(r,\,\varphi)$ independent of fiber orientation.
 In the following, we restrict
ourselves to radially symmetric
swelling functions $A(r)$; the neutral case of the flat disk is 
represented by $A(r) = 1$, $A(r)>1$ corresponds to local swelling,
$A(r)<1$ to local shrinking of fibers.
In order to calculate the change in metric by a swelling
function $A(r)$ we re-parametrize
the deformed shape using Gaussian normal coordinates.
The Gaussian radial coordinate is given by 
$\rho \equiv \int_0^r A(\tilde{r}) d\tilde{r}$, which is the
 distance of a point to the origin of the coordinate system following the
 surface and reduces to  the standard radial coordinate $r$ for a
 flat disk $A(r)=1$. The angular coordinate $\varphi$ remains 
 unchanged.
 In Gaussian coordinates a deformed fiber in radial direction has
 length $dl_\rho = d\rho = A(r) dr$, a deformed
 circumferential fiber 
 a length $dl_\varphi =  rA(r) d\varphi$.
 In general, the fiber length is related to the
 metric by $dl^2 = g_{\rho\rho} d\rho^2 + 2g_{\rho\varphi}d\rho d\varphi+
   g_{\varphi\varphi} d\varphi^2$ 
   such that the metric tensor of the swollen disk can be read
   off as 
	\begin{align}
	  \label{eq:metric_tensor_disk}
          \bar{\textbf{g}} = \begin{pmatrix}
			    1 & 0\\
			    0 & r^2(\rho)A^2(r(\rho))
                          \end{pmatrix}.
	\end{align}
in Gaussian normal coordinates.
This is the so-called target metric which represents the
preferred equilibrium state of the swollen disk \cite{Efrati2009}.
According to the Theorema Egregrium the Gaussian curvature $\bar{K}$ 
can be deduced solely from the metric tensor. Using the Brioschi formula
(with respect to the  Gaussian coordinates $\rho$ and $\varphi$
  of the metric) we find 
\begin{align}
  \bar{K}(\rho) &= -\frac{\partial_\rho^2(r(\rho)A(r(\rho)))}
                  {r(\rho)A(r(\rho))},
                  \label{eq:GaussKrho}
\end{align}
as a function of the Gaussian radial coordinate $\rho$.
  We can transform to the standard radial coordinate $r$ by using
  $dr/d\rho = 1/A(r)$, which gives
\begin{align}
 \bar{K}(r)  &= - \frac{A'(r) + r A''(r) - r A'^2(r)/A(r)}
               {r A^3(r)}.
  \label{eq:GaussK}
\end{align}
This is the Gaussian curvature if a shape with the metric
  (\ref{eq:metric_tensor_disk}) could be embedded
  into  three-dimensional Euclidian space.

\begin{figure}[t]
	\centering
  \includegraphics[width=0.9\linewidth]{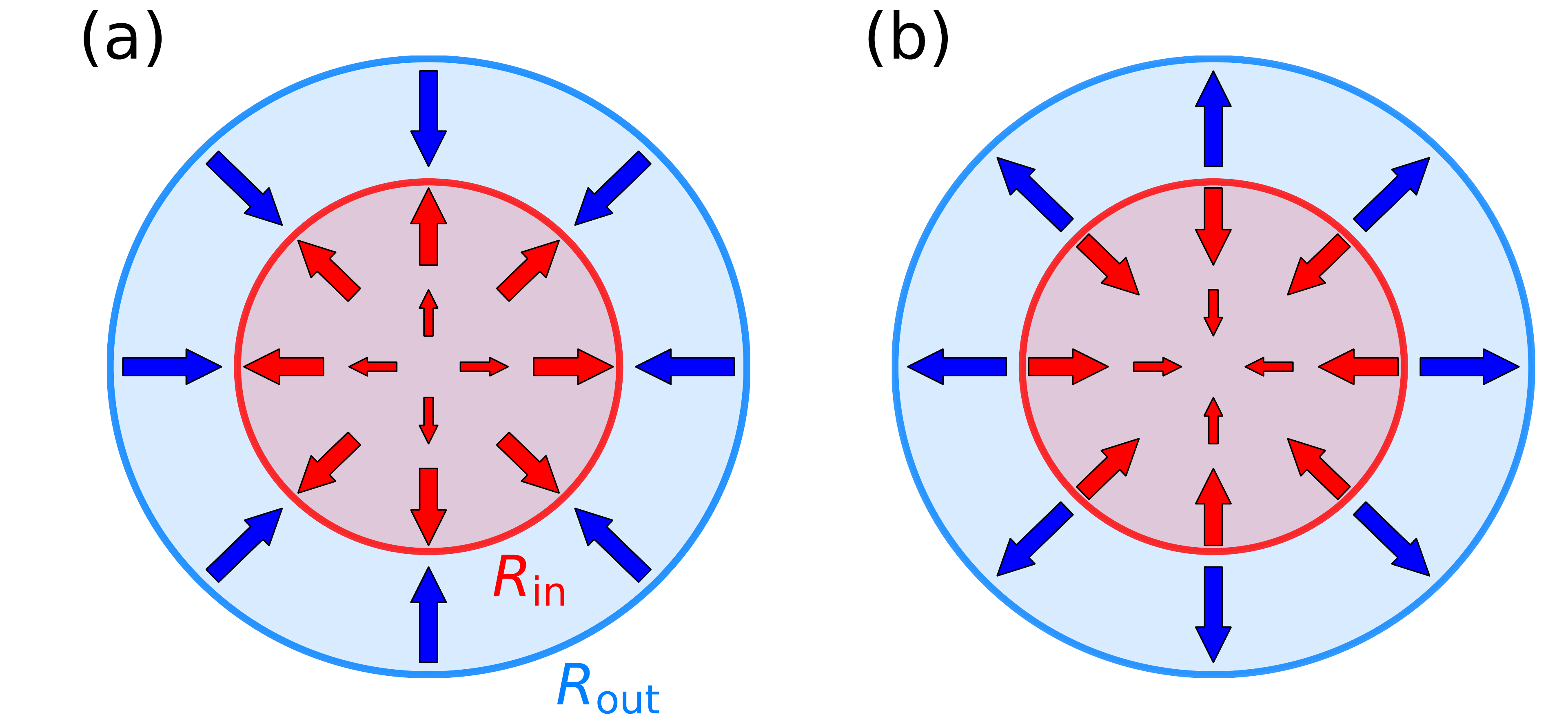}
    \caption{Schematic demonstration of the swelling pattern. The
                  disk is divided into two parts. The outer annulus (blue)
                  swells with a constant factor $\alpha$, the dinner disk
                  (red) with $1/\alpha$. (a): With $\alpha > 1$, the inner
                  disk expands and the annulus shrinks. (b): Shrinking in the
                  inner region and material expansion in the annulus with
                  $\alpha < 1$.}
		\label{fig:swell_sketch}
\end{figure}

In order to deform the disk into a curved shape, we define a simple
class of swelling patterns.
Similar to Pezulla {\it et al.} \cite{Pezulla2015}, we
  divide the disk into two parts: An inner disk with
  $0\le r \le R_{\mathrm{in}}$ and an outer annulus with
 $R_{\mathrm{in}}<r\le R_\mathrm{out}$. Within these two
 regions, the swelling function $A(r)$ shall be piecewise constant.
 To simplify things further, we define the inner disk to swell
 with a constant factor $\alpha$, while  the outer annulus
  shall always do the  exact opposite, i.e., it
 shrinks with the inverse constant factor $1/\alpha$.
 In total, the considered
 swelling functions $A(r)$ can  be written as
 \begin{align}
   \label{eq:def_alpha}
    A(r) =      \begin{cases}
      {\alpha}, & r\in [0, R_{\mathrm{in}}]\\
      \frac{1}{\alpha}, & r\in(R_{\mathrm{in}}, R_\mathrm{out}].
			\end{cases} 
\end{align}
The swelling process is thus defined by two simple control parameters: The
swelling factor $\alpha$ and the geometrical ratio
$R_\mathrm{in}/R_\mathrm{out}$, which will be kept constant at 
$R_\mathrm{in}/R_\mathrm{out}=0.5$ in the following 
so that we can focus on the influence of $\alpha$.
We can distinguish between two general cases: (a)  $\alpha>1$
means that the material in the  inner disk
is swelling, while the outer annulus is shrinking,
which is illustrated in Fig.\ \ref{fig:swell_sketch}(a).
The opposite case, (b) $\alpha < 1$, leads to a  shrinking inner
disk  and a swelling annulus (Fig.\ \ref{fig:swell_sketch}(b)).
In order to apply the  Brioschi formula  (\ref{eq:GaussK}),
  we have to smear out the step function at $r=R_{\mathrm{in}}$
  in eq.\ (\ref{eq:def_alpha}). For $\alpha>1$,
  we have  $A'(r)\approx 0$
  except around $r=R_{\mathrm{in}}$, where $-A'(r)$ is peaked.
  Therefore, the last term in the denominator in eq.\ (\ref{eq:GaussK})
  dominates and gives a positive Gaussian curvature
  which is peaked around  $r=R_{\mathrm{in}}$ if a surface with the
   piecewise  metric  \eqref{eq:metric_tensor_disk}
  could be embedded into three-dimensional space.
 Although the metric  \eqref{eq:metric_tensor_disk}
  is piecewise flat, Gaussian curvature has to be introduced
  because the inner disk is bonded to the outer annulus.
  This Gaussian curvature is positive  for $\alpha>1$ and negative
   for $\alpha<1$. 
   Because the metric can actually not be embedded, the Gaussian
   curvature will be redistributed on the entire disk to minimize
   the elastic energy. 
 We thus  expect a
target elliptic shape with an overall positive Gaussian curvature
for $\alpha>1$, and 
a hyperbolic target shape with a negative Gaussian curvature for $\alpha<1$.

Obviously, for $\alpha \neq 1$ the outer annulus
and the inner disk are incompatible at $r=R_\mathrm{in}$.
Therefore, the 
surface described by the metric \eqref{eq:metric_tensor_disk} has
no immersion in the Euclidean three-dimensional
embedding space and is an example of non-Euclidean
geometry \cite{Efrati2009}.
The actual shape of the surface in  three-dimensional
space is then defined by the minimization of the elastic energy,
where the elastic energy of the deformed
swollen state is defined with respect to the
above target metric $\bar{\textbf{g}}$
from  eq.\ \eqref{eq:metric_tensor_disk}.
The incompatibility of the two parts and the
resulting non-existence of an immersion means that there always is
 a residual elastic energy after minimization \cite{Efrati2009}.

The elastic energy contains a stretching and a bending
contribution.  
The stretching contribution is
caused by strains $\varepsilon_{ij} = (g_{ij} -\bar{g}_{ij})/2$, where
$\textbf{g}$ is the actual metric that the {\em deformed}
swollen state assumes, and given by \cite{Efrati2009}
\begin{align}
  E_{\mathrm{s}} &=  \int du_1 du_2 \sqrt{|\bar{\textbf{g}}|}  
            \frac{1}{8}  A^{ijkl} (g_{ij}-\bar{g}_{ij}) (g_{kl}-\bar{g}_{kl}),
                   \label{eq:Es}\\
  A^{ijkl} &= \frac{Y_\mathrm{2D}}{1-\nu^2} \left(
             \nu \bar{g}^{ij} \bar{g}^{kl}
             + \frac{1-\nu}{2} \left(\bar{g}^{ik} \bar{g}^{jl}+
             \bar{g}^{il} \bar{g}^{jk}
             \right) \right),
             \label{eq:A}
\end{align}
where we use  Einstein summation, raising of indices is
performed with the target metric, and  $(u_1,u_2)=(\rho,\varphi)$
in Gaussian  normal parametrization.
The elastic tensor $A^{ijkl}$  is given by 
the two dimensional Young modulus $Y_\mathrm{2D}$
and  the Poisson ratio $\nu$, which characterize the stretching elasticity
of the disk material.
The stretching energy thus penalizes deviations from the
target metric.

Likewise, the  bending energy is defined with the curvature
tensor $\textbf{L}$ and with respect to a target curvature tensor
$\bar{\textbf{L}}$, which represents a spontaneous curvature
of the material. We assume that local isotropic swelling
does not introduce any spontaneous curvature to the system
such that $\bar{\textbf{L}}=0$.
The general expression for the   bending energy is \cite{Efrati2009}
\begin{align}
  E_\mathrm{B} &= \int du_1 du_2 \sqrt{|\bar{\textbf{g}}|}
             \frac{h^2}{24} A^{ijkl} \left( L_{ij}-\bar{L}_{ij}\right) 
                 \left( L_{kl}-\bar{L}_{kl}\right)
                 \label{eq:EB}\\
               &\approx \int d\bar{A} \frac{1}{2}\kappa_\mathrm{B}
  \left( 4H^2 - 2(1-\nu) K \right),
  \label{eq:EB_HK}
\end{align}
where the last line applies to  $\bar{\textbf{L}}=0$, 
$H$ is the mean curvature, $K$ the Gaussian curvature,
and  $\kappa_\mathrm{B} = Y_\mathrm{2D}h^2/(12(1-\nu^2))$ 
the bending modulus of the disk.
The bending energy penalizes deviations from the flat
shape for vanishing target curvature  $\bar{\textbf{L}}=0$.
The last line in (\ref{eq:EB_HK}) is an approximation because
we assume $\bar{g}_{ij}\approx g_{ij}$ in $A^{ijkl}$.
Typical strains $\varepsilon_{ij}$
are $\propto (1-\alpha)^2$ such that corrections are
${O}((1-\alpha)^2E_\mathrm{B})$
and will be small at the transition for thin disks
(see eq.\ (\ref{eq:alphace}) and Fig.\ \ref{fig:alphaFvK} below).

For numerical energy minimization the disk and its elastic energies
\eqref{eq:Es} and \eqref{eq:EB} have to be  suitably discretized.

\subsubsection{Model}

We calculate the disk's shape with the help of a numerical energy
minimization and  use a simple spring mesh model for discretization.
The disk is
triangulated with a Delaunay triangulation (implemented with the fade2D
library \cite{Fade2D}), where every edge $i$ between two vertices represents a
mechanical spring with a rest length $l_i$. The fineness of the mesh is
controlled by the number of vertices $n_\mathrm{B}$ on the boundary of the
disk. In this model, a swelling process is performed by a simple
multiplication of the springs' rest lengths with the swelling function
$A(r)$. The discretized  version of the
elastic stretching energy \eqref{eq:Es}
can be written as the sum over all
spring energies,
\begin{align}
      	\label{eq:stretch_energy_spring}
  E_{\mathrm{s}} = \sum\limits_i\frac{1}{2}k_i
  (|\vec{r}_{2,i} - \vec{r}_{1,i}| - l_i)^2.
 \end{align}
 The vectors $\vec{r}_{2,i}$ and $\vec{r}_{1,i}$ describe the positions of the
 vertices that define the beginning and the end of a spring. The spring
 constants are denoted by $k_i$. In a hexagonal mesh,
 the two-dimensional Young modulus $Y_\mathrm{2D}$ is given
 by the spring constant $k$ and the Poisson ratio
 $\nu$ is fixed, 
 \cite{Ostoja-Starzewski2002,Nelson1988}
\begin{align}
	Y_\mathrm{2D} = \frac{2}{\sqrt{3}}k ~~~\mbox{and}~~\nu = \frac{1}{3}.
\end{align}

\begin{figure}[b]
		\centering
		\includegraphics[width=0.5\linewidth]{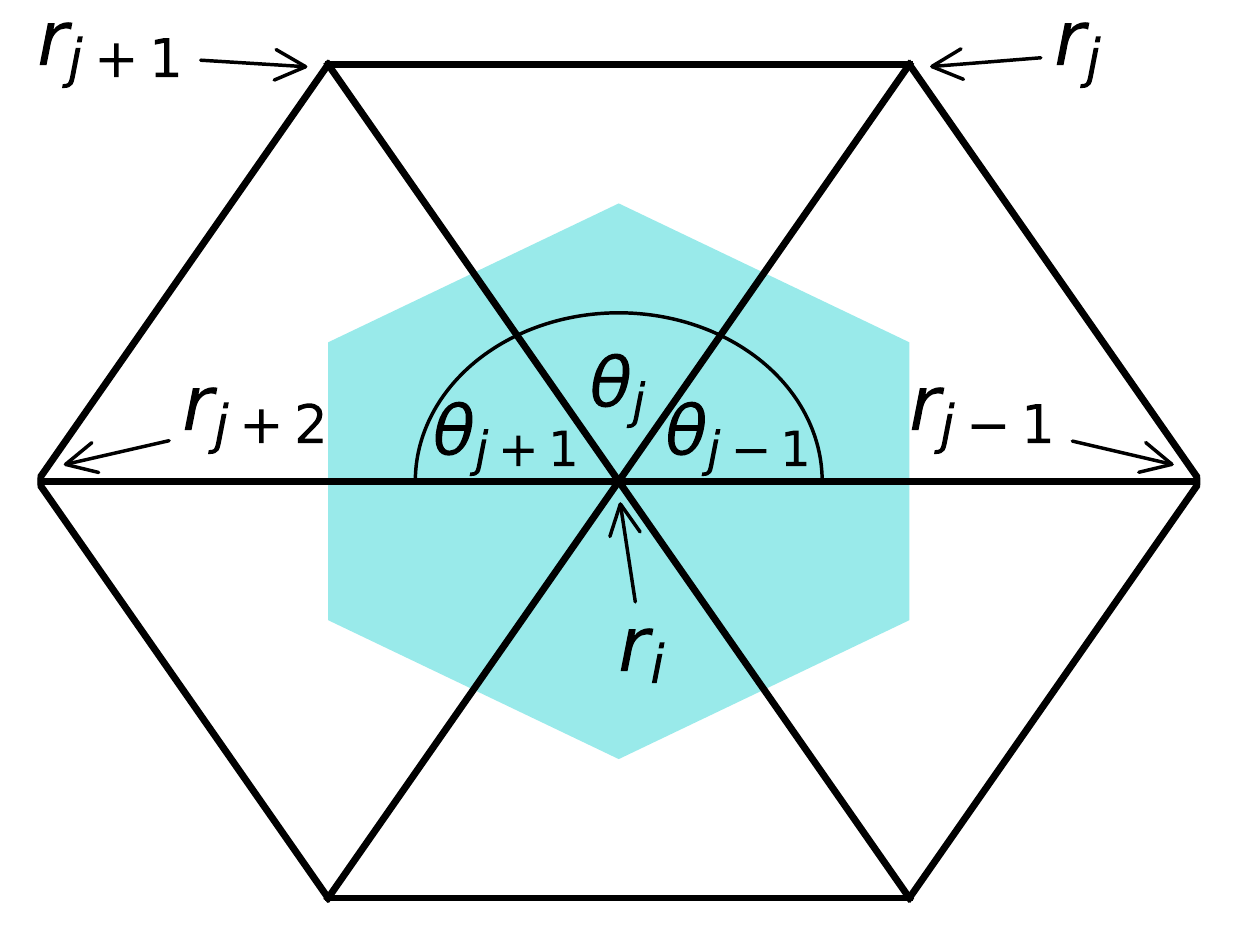}
		\caption{Illustration of the direct neighborhood of a vertex
                  at position $\vec{r}_i$. The area $A_i$ that is associated
                  with this vertex is shown cyan. The angle between the
                  springs to the neighbor vertices at $\vec{r}_j$ and
                  $\vec{r}_{j+1}$ is called $\theta_j$ and is used in the
                  calculation of the Gaussian curvature.}
		\label{fig:associated_area}
\end{figure}

In order to evaluate the bending energy \eqref{eq:EB_HK}  on
the spring mesh, the curvatures $H$ and $K$ have to be calculated
on the mesh. The mean curvature $H_i$ at a mesh vertex $i$
can be expressed in terms of an area gradient \cite{SurfaceEvolver}:
\begin{align}
	H_i = \frac{3}{2}\frac{|\nabla_i A_i|}{A_i}.
\end{align}
The quantity $A_i$ represents the area in the mesh that is associated to the
vertex $i$, see the colored area in Fig.\ \ref{fig:associated_area}. The
gradient $\nabla_i A_i$ then describes derivatives of this area with respect
to the coordinates of the vertex $i$.
The Gaussian curvature $K$, on the other hand, can be calculated using the
Gauss-Bonnet-theorem. We find
\begin{align}
	K_i=(2\pi-\sum\limits_j \theta_j )/(A_i/3)
\end{align}
where  $\theta_j$ is the angle  between the neighboring
vertices $j$ and $j+1$ of the vertex $i$
located at $\vec{r}_i$, see Fig.\ \ref{fig:associated_area}.
Finally, the discretized bending energy \eqref{eq:EB_HK} becomes 
\begin{align}
      	\label{eq:EB_HK_discrete}
   E_\mathrm{B} = \sum\limits_i \frac{A_i}{3}  \kappa_\mathrm{B}
  \left( 2H_i^2 - \frac{2}{3} K_i \right).
 \end{align}

\begin{figure}[t]
	\centering
		\includegraphics[width=0.90\linewidth]{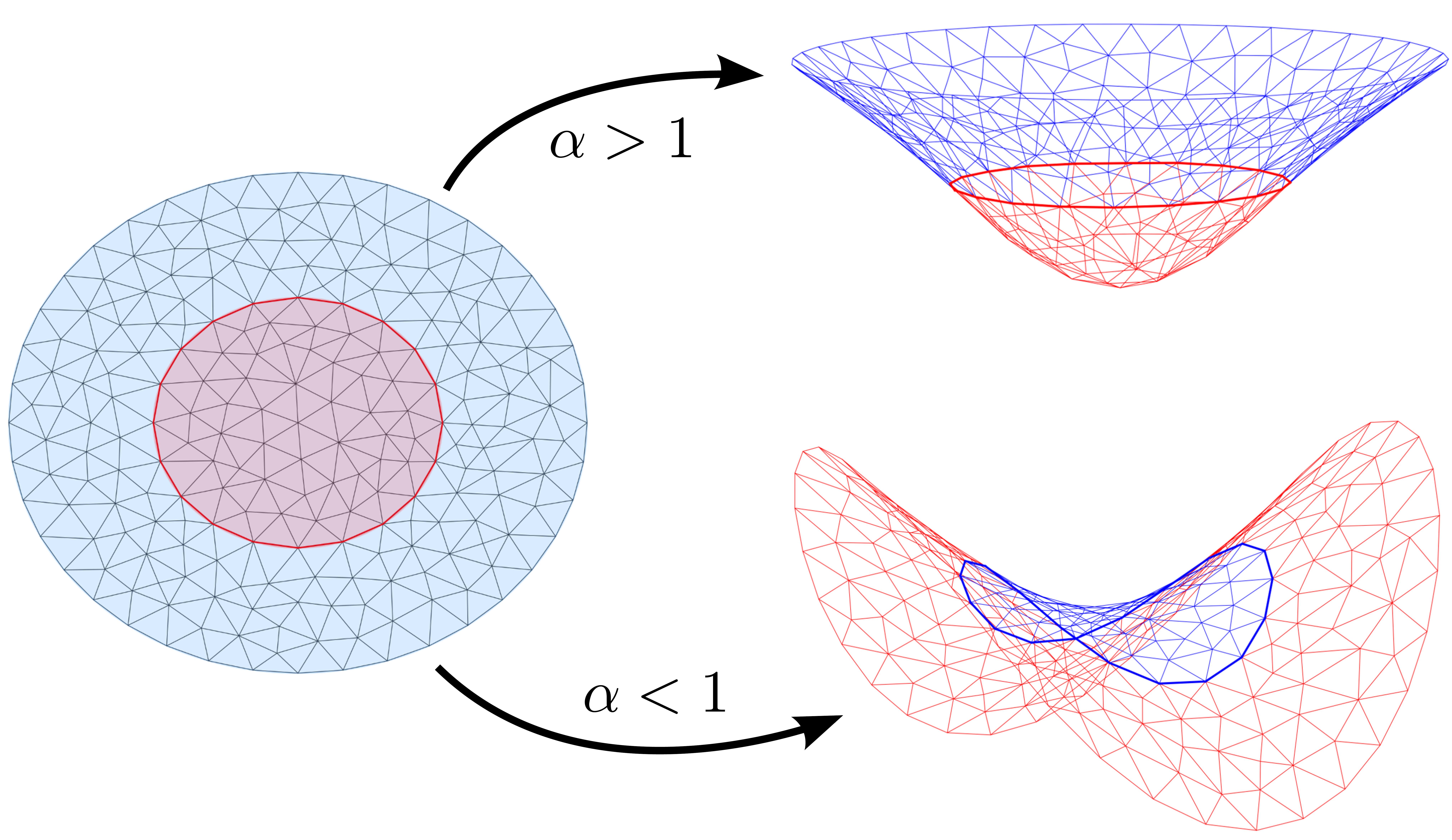}
	\caption{Example Delaunay triangulation with $n_\mathrm{B} =
                  40$ vertices on the boundary representing the spring
                  mesh. Left side: flat disk with $\alpha = 1$. Right side:
                  resulting elliptic shape for $\alpha > 1$
                  (swelling of interior disk) and hyperbolic
                  shape for $\alpha < 1$ (shrinking of interior disk).
                  Swollen springs are shown in red,
                  while shrunk springs are shown in blue.}
		\label{fig:meshes}
\end{figure}

The total energy $E=E_{\mathrm{s}}+ E_\mathrm{B}$  has to be minimized
with respect to all vertex coordinates in the three-dimensional embedding
space. In
order  to overcome possible local energy minima, small fluctuations can be
added to the vertex coordinates in terms of a random displacement
$\vec{r}_i\rightarrow \vec{r}_i+\vec{\delta}_i$ with $|\vec{\delta}_i|\ll
l$. After minimizing  the global energy minimum with respect
to all vertex positions, the resulting mesh 
represents the preferred configuration of the swollen and
deformed disk, see the illustration in Fig.\ \ref{fig:meshes}.

%

\subsubsection{Control parameters}

After all, our system of the swelling elastic disk is defined by a small set
of dimensionless control parameters. These are the previously
mentioned swelling
factor $\alpha$ and the ratio of the inner and outer radius
$R_\mathrm{in}/R_\mathrm{out}$. In addition,
we also want to be able to describe a disk where the
inner disk and the outer annulus consist of
different materials \cite{Pezulla2015}.
Therefore, we introduce
 different elastic moduli and thus different spring constants. The spring
 constant $k_\mathrm{in}$ is valid for interior springs
 with $r \leq R_\mathrm{in}$,
 while $k_\mathrm{out}$ belongs to outer springs with
 $r> R_\mathrm{in}$, and  the ratio
$k_\mathrm{in}/k_\mathrm{out}$ is another control parameter. Finally, the
thickness of the disk has an influence, even in a two-dimensional model:
the relative importance of the bending energy   \eqref{eq:EB_HK}
is governed by  $\kappa_\mathrm{B}/Y_\mathrm{2D} \propto h^2$, i.e.,
a thicker disk is harder to bend.
This is usually captured by a dimensionless
F\"oppl-von K\'arm\'an number, a dimensionless ratio of Young modulus and
bending modulus, which we define for our disk as 
\begin{align}
  \gamma_\mathrm{FvK} \equiv
  \frac{Y_\mathrm{2D} R_\mathrm{out}^2}{\kappa_\mathrm{B}}
  = 12(1-\nu^2)\frac{R_\mathrm{out}^2}{h^2}.
  \label{eq:FvK}
\end{align}
The F\"oppl-von K\'arm\'an number is large for thin  disks and
 is the fourth and last control parameter of our system.

\subsection{Results}

\begin{figure*}[t!]
	\centering
		\includegraphics[width=1.0\linewidth]{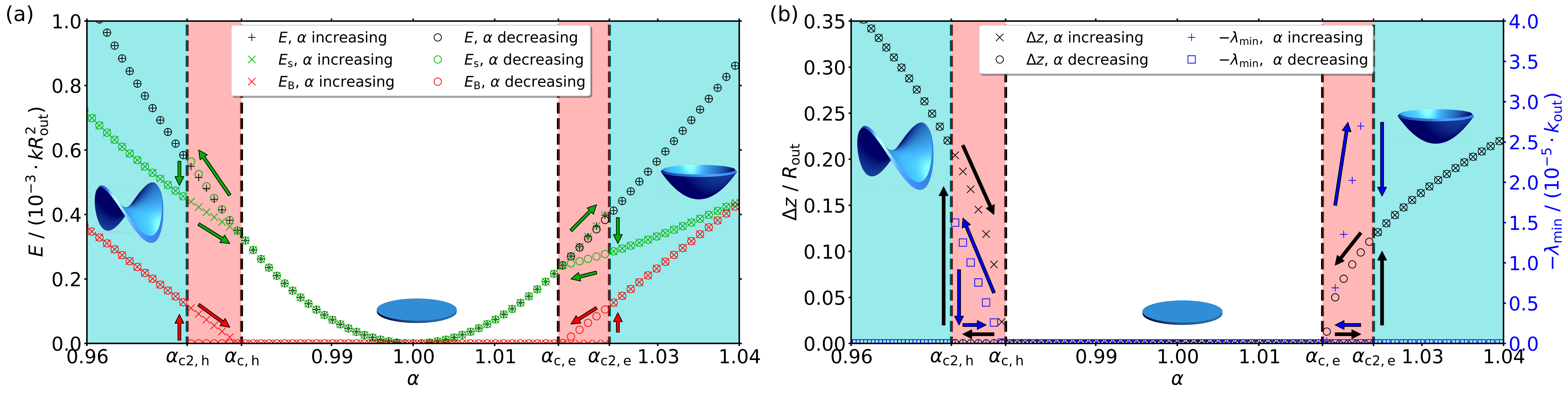}
                \caption{Energies (a), shape's height $\Delta z$ and 
                   negative
                   lowest Hessian eigenvalue $-\lambda_\mathrm{min}$ (b)
                  in the spring mesh model as functions of the stretch factor
                  $\alpha$.  Circles denote numerical values calculated with
                  decreasing $\alpha$, while crosses are related to increasing
                  $\alpha$.  The disk is always flat if it is located in the
                  white area and always curved in the cyan regions. The blue
                  shapes illustrate the corresponding conformations of the
                  disk.  The red areas mark the regions of pseudo-hysteretic
                  effects. Arrows illustrate the directions inside the
                  hysteresis loops.  The simulated disk had a mesh with
                  $n_\mathrm{B}=120$ boundary vertices,
                  $R_\mathrm{in}/R_\mathrm{out} = 0.5$,
                  $\gamma_\mathrm{FvK} = 600$,
                  $k_\mathrm{in}/k_\mathrm{out} = 0.24$ and maximum
                  fluctuations of
                  $\delta_\mathrm{max}= 5\times 10^{-4}R_\mathrm{out}$.}
		\label{fig:energyplot}
\end{figure*}

Starting with a flat disk with $\alpha = 1$, we increase/decrease
  $\alpha$ in small steps $\Delta \alpha$ and minimize the energy after
  each step. Figure \ref{fig:energyplot}(a) shows the resulting energies:
  the  total energy, the spring energy and the bending energy (separated in
  mean curvature  and  Gaussian curvature part), as functions
  of $\alpha$, while Fig.\ \ref{fig:energyplot}(b) shows the total
  height $\Delta z$ of the shape. The shape can  deform  both
    into positive and negative $z$-direction with equal probability;
    we count the height  $\Delta z$ of the shape always as the positive
  absolute value of the maximal difference of $z$-coordinates. 
    For small changes of $\alpha$ the disk
  stays flat at first, only the spring energy increases quadratically
  because of  the change of the springs' rest lengths. We have
  $E=E_\mathrm{s}$ in this regime.
  Swelling ($\alpha>1$) or shrinking ($\alpha<1$) of the interior
    disk imparts elastic compression or stretching energy to the
    flat state, which is released in the snapping transition.
  At a critical swelling factor
  $\alpha_{\mathrm{c2,e}}$ for increasing $\alpha$ (or
  $\alpha_{\mathrm{c2,h}}$ for decreasing $\alpha$, respectively)
  a transition into a curved conformation  with $\Delta z >
  0$ occurs.
  We find two stable curved configurations: for increasing $\alpha$
   above   $\alpha_{\mathrm{c2,e}}>1$
   the disk snaps into an elliptic (subscript ``e'') dome-like shape, 
   while
   it snaps into 
   a hyperbolic (subscript ``h'') saddle  for decreasing $\alpha$ beyond
   $\alpha_{\mathrm{c2,h}}<1$ (see Fig.\ \ref{fig:meshes}).
  At these transitions,  $E_\mathrm{s}$ is reduced,
  because the springs can relax to
  a certain degree. On the other hand, $E_\mathrm{B}$ is increased because
  of the increased curvatures in the dome- or saddle-like
   shapes. Increasing (decreasing) $\alpha$ again
  in order to get back to  $\alpha = 1$,
  we do not see a   transition back  into the flat state
  at $\alpha_{\mathrm{c2,e}}$ (or
  $\alpha_{\mathrm{c2,h}}$, respectively). Instead, the shape remains
   curved for $\alpha <\alpha_{\mathrm{c2,e}}$
  ($\alpha > \alpha_{\mathrm{c2,h}}$). In the following, the curved
   disk flattens continuously with $E_\mathrm{B}$ and $\Delta z$ decreasing
    until $\alpha = \alpha_\mathrm{c,e}$ (or
  $\alpha = \alpha_\mathrm{c,h}$) is reached. There, $E_\mathrm{B}$ and
   $\Delta z$ vanish continuously, and the disk is flat
   again. In conclusion, we find an apparent  hysteresis loop in the
   deformation behavior within the red areas between $\alpha_\mathrm{c,e}$ and
$\alpha_\mathrm{c2,e}$ (or $\alpha_\mathrm{c,h}$ and
$\alpha_\mathrm{c2,h}$).

\subsubsection{Pseudo-hysteresis and long-wavelength bifurcation}

The stability of the disk's conformation upon approaching the transition
can be analyzed in more detail with the help of the
eigenvalues of the Hessian matrix of the system's total energy. If the smallest
eigenvalue $\lambda_\mathrm{min}$ becomes negative,  there is a 
deformation mode leading directly
to a lower energy, and the system becomes unstable. The
blue scale  on the right side of Fig.\ \ref{fig:energyplot}(b) shows the
smallest eigenvalue of our elastic system (please note that we show the
negative eigenvalue $-\lambda_\mathrm{min}$).
It is zero in the flat configuration
until $\alpha$ exceeds $\alpha_\mathrm{c,e}$ (or $\alpha_\mathrm{c,h}$).
Then,
still in the flat configuration in the red area, $\lambda_\mathrm{min}$
becomes significantly negative indicating that the system is unstable.
This means
that already in the entire red area, there is
an unstable  deformation mode available that
leads directly into the curved
conformation. After the transition, the curved conformation remains
stable.  Therefore, we conclude  that the red area is {\it not} an area of
genuine hysteresis. The transition to the curved shape could directly happen at
$\alpha_\mathrm{c,e}$ (or $\alpha_\mathrm{c,h}$) if the system finds the
existing unstable deformation mode. The disk remains flat in the red area
only for numerical
reasons, and the values of
$\alpha_\mathrm{c2,e}$ and $\alpha_\mathrm{c2,h}$ and, thus,
the size of the red region 
actually shrinks  if we increase  random displacements 
$|\delta_i|$ (numerical fluctuations)
that are imposed. In the experimental system,
we expect that thermal fluctuations will always allow the
disk to find the unstable mode such that hysteresis
will be absent.

\begin{figure}[t!]
	\centering
		\includegraphics[width=0.9\linewidth]{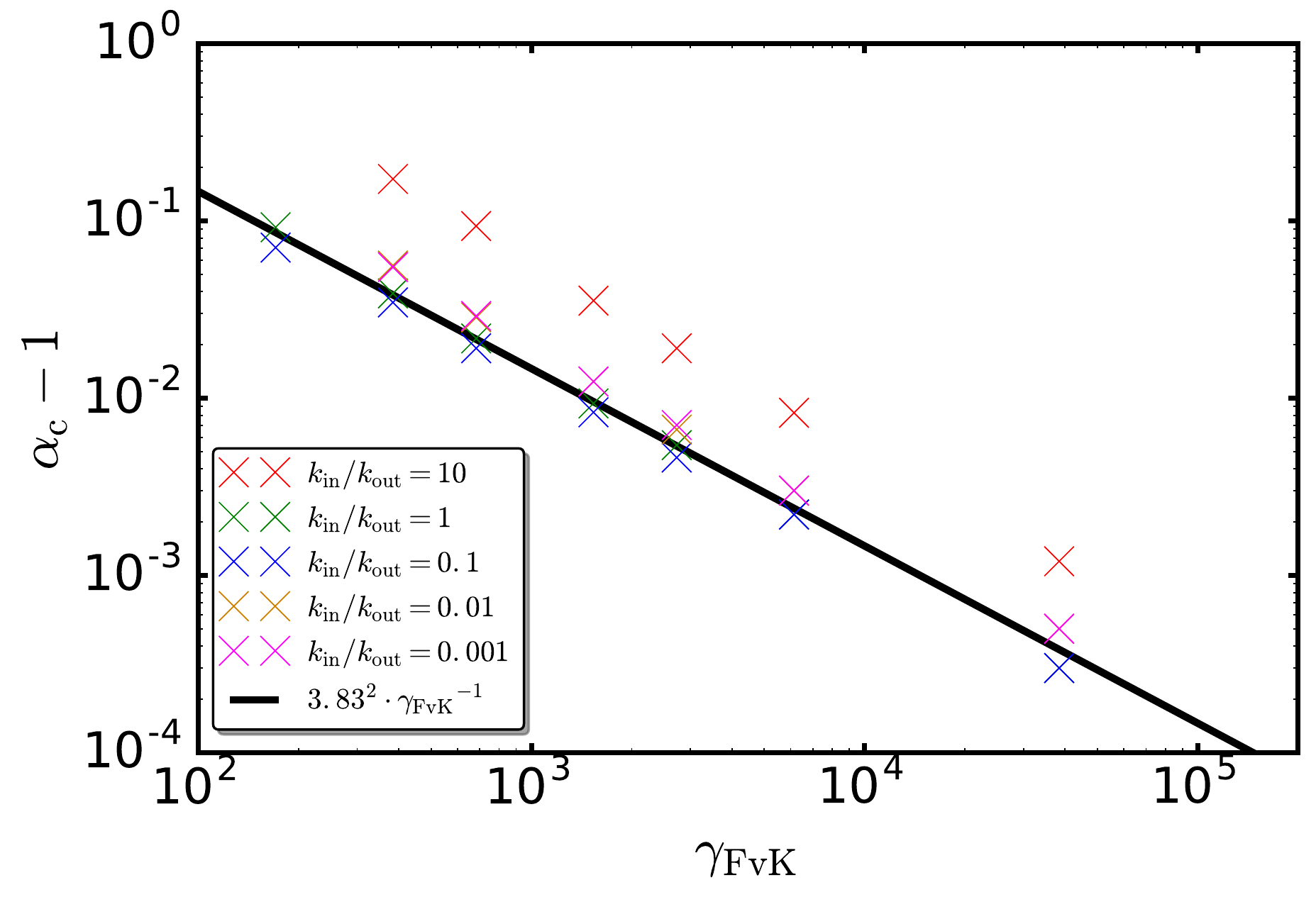}
                \caption{Critical swelling factor $\alpha_{c,e}$ as a function
                  of $\gamma_{\mathrm{FvK}}$ for
                  $R_\mathrm{in}/R_\mathrm{out} = 0.5$ and different values of
                  $k_\mathrm{in}/k_\mathrm{out}$. The solid black line
                  represents the theory curve given by
                  eq.\ \eqref{eq:alphace}.}
		\label{fig:alphaFvK}
\end{figure}

We can perform a linear stability analysis of the flat compressed state
of the inner disk 
 in order to further characterize the bifurcation into an  elliptic dome
 for increasing $\alpha$ above $\alpha_\mathrm{c,e}$.
 In the limit of a small stiff outer annulus
($k_\mathrm{in}/k_\mathrm{out}\ll 1$
 and $R_\mathrm{in}/R_\mathrm{out} \approx 1$), 
 the effect of swelling the
 interior with a factor $\alpha>1$ is to establish a compressive
 homogeneous pre-stress
 $\sigma_{xx} = \sigma_{yy} = -\sigma_0 = -Y_\mathrm{2D} (\alpha-1)$
  in the interior.
We can perform 
a linear stability analysis of the flat state $z(x,y)=0$ of an infinite
plate under pre-stress $-\sigma_0$ using  plate theory. Expanding
the Airy stress function 
$\chi(x,y) = -\sigma_0(x^2+y^2)/2+ \chi_1(x,y)$
and  the normal displacement $z(x,y) = z_1(x,y)$ around
the flat, homogeneously pre-stressed state we find the following
plate equations to linear order in $\chi_1$ and $z_1$:
\begin{equation}
  \kappa_\mathrm{B} \nabla^4 z_1 +\sigma_0 \nabla^2 z_1=0~,~~
  \nabla^4 \chi_1 =0.
\label{eq:stab}
\end{equation}
An Ansatz $\chi_1 = a e^{i\vec{q}\cdot\vec{r}}$ and
$z_1 = b  e^{i\vec{q}\cdot\vec{r}}$ for an oscillatory
instability of the flat state with a two-dimensional
wave vector $\vec{q} = (q_x,q_y)$ leads to the condition 
\begin{equation}
   \sigma_0 = Y_\mathrm{2D}(\alpha-1) = \kappa_\mathrm{B} q^2,
\end{equation}
  which  is fulfilled for $\alpha> \alpha_\mathrm{c,e}$ with
  $\alpha_{c,e}-1 \approx \kappa_\mathrm{B} q^2/Y_\mathrm{2D}$.
  The resulting instability
  is a long-wavelength instability, i.e.,
  sets in at the smallest available wave vector $q$,
   as opposed to buckling of a
  spherical shell under pressure, where the pressure also leads to a
  homogeneous compressive pre-stress, but the buckling instability
  is a short-wavelength instability
  because of  the non-vanishing background curvature
  \cite{Hutchinson1967,Baumgarten2018,Baumgarten2019}.
  For an inner disk of radius $R_\mathrm{in}\approx
  R_\mathrm{out}$ the shortest available wave vectors
  have  $q  \sim 1/R_\mathrm{out}$.
 Closer inspection shows that the unstable radially symmetric, oscillating
  modes are
  $z_1(r) = b J_0(r\sqrt{\sigma_0/\kappa_\mathrm{B}})$ (with
 the Bessel function $J_0$).
The approximate boundary condition  $\partial_r z_1(R_\mathrm{out})=0$
for a small stiff outer annulus  leads to 
\begin{equation}
    \alpha_{c,e}-1 \approx
    3.83^2  \frac{\kappa_\mathrm{B}}{Y_\mathrm{2D}R_\mathrm{out}^2} =
    3.83^2  \gamma_\mathrm{FvK}^{-1},
    \label{eq:alphace}
\end{equation}
where $3.83$ is the first zero of the Bessel function $J_1(x)$.
This is in good agreement with numerical results even
    for $k_\mathrm{in}/k_\mathrm{out}< 1$
    and $R_\mathrm{in}/R_\mathrm{out} =0.5$
 as shown in Fig.\ \ref{fig:alphaFvK}.

  The stability analysis with the stability equation  \eqref{eq:stab}
  also  shows  that a genuine hysteresis  should
  be absent, and the bifurcation is a supercritical pitchfork
  bifurcation similar to the Euler buckling bifurcation of a beam.
  This is in contrast 
   to buckling of a spherical shell under pressure,
  which is a subcritical bifurcation
  \cite{Baumgarten2019}.

\section{Re-establishing  hysteresis}
 
\subsection{Framing the disk and additional attractive interaction}

Now we want  to modify the system in a way that a genuine hysteresis
is re-established,
which will also be present in experiments. The basic idea is to energetically
penalize slightly deformed intermediate states of the disk
during the transition to the curved shape
resulting in an  additional energy barrier for  this transition.
This barrier has to be
overcome or decreased by additional swelling before the disk can snap
into a curved state and stabilizes the flat disk.

\begin{figure}[b]
		\centering
		\includegraphics[width=0.7\linewidth]{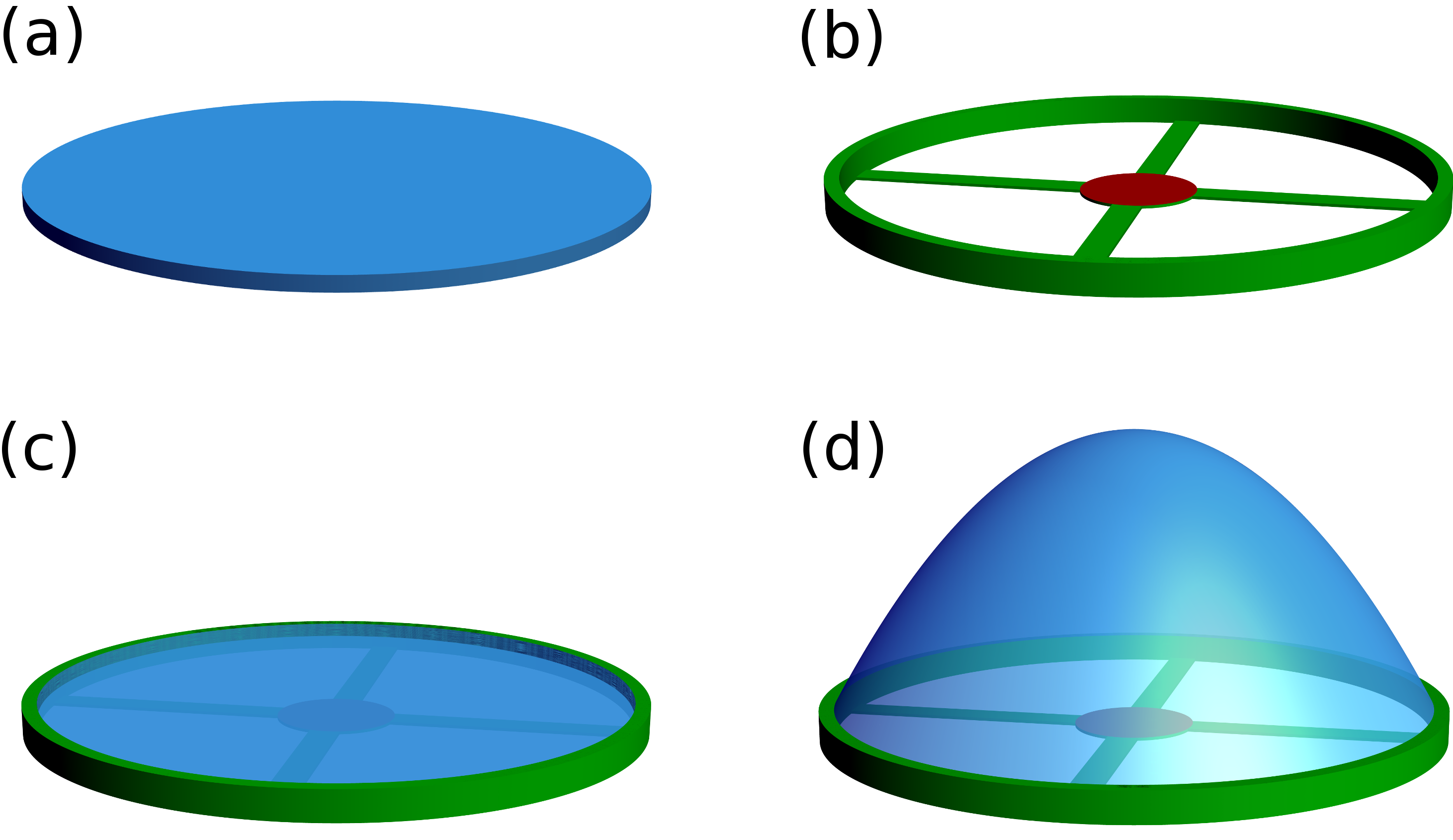}
		\caption{Illustration of the set-up of a flat disk inside of a
                  fixed frame. The flat disk (a) is placed inside a fixed and
                  undeformable frame (b)
                   with an attractive central  region
                  (in red). Disk and frame are compatible in the
                  flat state (c). If the disk swells uniformly, an elliptic
                  dome-like shape results (d).}
		\label{fig:frame_sketch}
\end{figure}

In order
to realize that, the first step is to ``frame the disk'':
we combine our flat disk with
radius $R_\mathrm{out}$ in the $xy$-plane (Fig.\ \ref{fig:frame_sketch}(a))
with an additional fixed, undeformable frame (Fig.\ \ref{fig:frame_sketch}
(b)). As a result, the boundary of the disk is now fixed (Fig.\
\ref{fig:frame_sketch}(c)). Therefore, the piecewise constant swelling
function \eqref{eq:def_alpha} is replaced by a 
globally constant, homogeneous swelling factor $\alpha$ for the whole disk
(but not for the frame);
 $\alpha>1$ still corresponds to swelling the interior
of the disk and,  accordingly, leads to a transition
into
a dome-like shape (Fig.\ \ref{fig:frame_sketch}(d)). A saddle shape
can no longer be realized in this set-up.
Framing is equivalent to the above limit of a very thin and stiff
outer annulus.

%


The second step in order to create a genuine
hysteresis is to introduce  an additional attractive
interaction between the frame and the disk.
The central region of the frame $A_c$ 
(red area in Fig.\ \ref{fig:frame_sketch}) attracts the disk leading to an
additional potential energy for the disk. Inspired by an attractive van der
Waals force, we choose the attractive part of a Lennard-Jones potential for
the interaction,	
\begin{align}
\label{eq:LJ-pot}
	v_\mathrm{pot}(z) = 
  4\varepsilon\left[
  \left(\frac{z+\sqrt[6]{2}\sigma}{\sigma}\right)^{-12}
   -\left(\frac{z+\sqrt[6]{2}\sigma}{\sigma}\right)^{-6}\right],
\end{align}
with the total attractive potential energy $E_\mathrm{pot} = \int_{A_c} dA
v_\mathrm{pot}(z)$ or $E_\mathrm{pot} = \sum_{i\in A_c} A_i
v_\mathrm{pot}(z_i)$ for the  mesh model of the  disk.
The attractive potential has a finite range $\sigma$, and a potential
  depth $-\varepsilon$ at $z=0$;
the force of this potential vanishes in the completely flat state ($z=0$)
improving the numerical stability. For
$\sigma \ll R_\mathrm{out}$, on the other hand,
the force nearly vanishes also in the
completely deformed state
because most parts of the disk are out of the potential
range. In conclusion, only the transition itself is energetically
penalized.

\subsection{Hysteresis and short-wavelength bifurcation}

In order to gain insight into the influence of 
an attractive potential $v_\mathrm{pot}(z)$ onto the
instability of the swelling disk, we can consider the
case where the attractive potential  acts over the
whole area of the disk, i.e., $A_c=A$. 
Then the linear stability analysis leads to a short-wavelength
instability as eq.\ \eqref{eq:stab} becomes modified to
\begin{equation}
  \kappa_\mathrm{B} \nabla^4 z_1 +\sigma_0 \nabla^2 z_1
  + v_\mathrm{pot}''(0)z_1=0
\label{eq:stab2}
\end{equation}
resulting in an instability condition $\kappa_\mathrm{B} q^4-\sigma_0q^2 +
v_\mathrm{pot}''(0)<0$ (if $v_\mathrm{pot}'(0)=0$ and
with $v_\mathrm{pot}''(0)= 36\times 2^{2/3} \varepsilon/\sigma^2$ for the
potential \eqref{eq:LJ-pot}).
This is exactly equivalent to the wrinkling condition of a
membrane on an elastic substrate (or a Winkler foundation) under compressive
stress \cite{Huang2005}, where $v_\mathrm{pot}''(0)$ corresponds to the
substrate stiffness. 
Interestingly, this is also  equivalent to the short-wavelength
instability condition for buckling of a
  spherical shell under pressure with  the 
  homogeneous compressive pre-stress playing the role of the pre-stress from
  homogeneous pressure and the curvature of the potential
  $v_\mathrm{pot}''(0)$ playing the
  role of the background curvature term   \cite{Baumgarten2018}.
Now,  an instability sets in at the smallest $\sigma_0$,
  for which the instability condition can by fulfilled, which is for
  $\sigma_0> 2\sqrt{\kappa_\mathrm{B} v_\mathrm{pot}''(0)}$ or for
  $\alpha> \alpha_\mathrm{c,f}$ with
$\alpha_{c,f}-1  = {2\sqrt{\kappa_\mathrm{B}
      v_\mathrm{pot}''(0)}}/{Y_\mathrm{2D}}\propto h^2$
(subscript ``f'' for framed disk)
  and at  the wave vector
$q_0  = \left( {\sigma_0}/{2\kappa_B} \right)^{1/2} =
(v_\mathrm{pot}''(0)/\kappa_\mathrm{B})^{1/4}$. This 
 is a short-wavelength instability with  $q_0>1/L$ if
 $v_\mathrm{pot}''(0)$ is sufficiently large.
 We also expect to find a subcritical bifurcation
with hysteresis in analogy 
   to buckling of a spherical shell under pressure
  \cite{Baumgarten2019}.

 For a localized potential, i.e., if the attractive region $A_c$ is smaller
 than $A$ as  in Fig.\ \ref{fig:frame_sketch}, we expect that
 the  critical swelling factor is further increased such that the
 unstable wavelength 
 $1/q_0 = \left( {2\kappa_B}/\sigma_0  \right)^{1/2}$ fits  into the
 size $\sqrt{A_c}$ of the attractive region.
 This
  results in a condition  $\sigma_0 > \max[2\kappa_\mathrm{B}/A_c, 2\sqrt{\kappa_\mathrm{B} v_\mathrm{pot}''(0)}]$.

  \begin{figure*}[t!]
	\centering
		\includegraphics[width=1.0\linewidth]{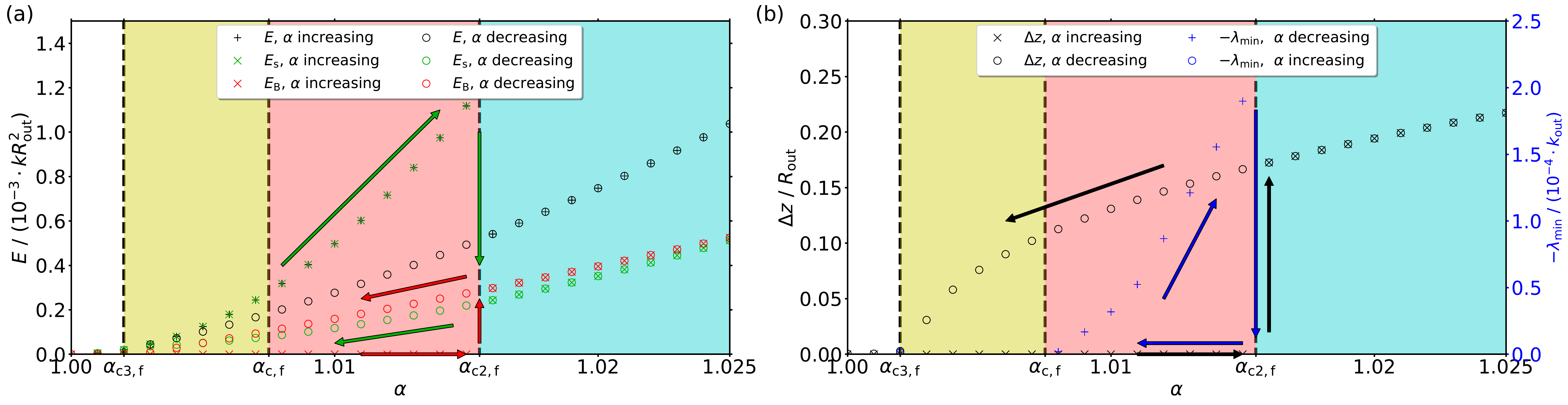}
		\caption{Energies (a), shape's height $\Delta z$ and negative
                   lowest Hessian eigenvalue $-\lambda_\mathrm{min}$ (b)
                  in the spring mesh of a disk in a fixed frame with
                  attractive potential as functions of the swelling factor
                  $\alpha$.  Crosses denote numerical values calculated with
                  increasing $\alpha$, while circles are related to decreasing
                  $\alpha$.  The region with a genuine hysteresis is marked in
                  yellow, while pseudo-hysteretic effects are again marked in
                  red. The disk is always flat if it is located in the white
                  area and always curved in the cyan region.  Arrows
                  illustrate the directions inside the hysteresis loops.  The
                  simulated system is the same system from
                  Fig.\ \ref{fig:energyplot} but with an additional potential
                  energy for each vertex given by eq.\ \eqref{eq:LJ-pot}.  The
                  simulated disk had a mesh with $n_\mathrm{B}=120$ boundary
                  vertices,
                  $\gamma_\mathrm{FvK} = 1066$ and maximum fluctuations of
                  $\delta_\mathrm{max}=  10^{-4}R_\mathrm{out}$.  The
                  parameters of the attractive potential were set to
                  $\varepsilon =
                  7.3\times 10^{-7}\,k_\mathrm{in}$ and
                  $\sigma = 0.01\,R_\mathrm{out}$.
                  The potential acted in an
                  inner region $A_c$ with a radius of $0.2\,R_\mathrm{out}$.}
		\label{fig:energies_fixedFrame}
\end{figure*}

Analogously  to Fig.\ \ref{fig:energyplot}, the behavior of the
system including frame and attractive interaction  is shown in Fig.\
\ref{fig:energies_fixedFrame}.
Increasing $\alpha$ starting at $\alpha = 1$,
the framed
disk again stays flat at first until $\alpha_\mathrm{c2,f}$ is reached,
where the transition into the curved, dome-like shape occurs. There, we see a
significant reduction of the spring energy and an increase of the bending
energy. In contrast to Fig.\ \ref{fig:energyplot}, also the total energy is
reduced drastically during the transition. Decreasing $\alpha$ again, the
behavior is qualitatively the same as before, the shape continuously
flattens but
stays curved until $\alpha_\mathrm{c3,f}$ is reached, where the disk is flat
again. The significant difference to the simple set-up from above can be found
by taking a look at the smallest Hessian eigenvalue $\lambda_\mathrm{min}$
(Fig.\ \ref{fig:energies_fixedFrame}(b), blue scale). Between
$\alpha_\mathrm{c3,f}$ and $\alpha_{c,f}$ (yellow area), there are no negative
eigenvalues, which  means that both the flat disk and the curved shape are
(meta-)stable in this region.  Only if $\alpha$ exceeds $\alpha_\mathrm{c,f}$,
$\lambda_\mathrm{min}$ becomes negative signalling an unstable flat shape. In
conclusion, the region between $\alpha_\mathrm{c,f}$ and
$\alpha_\mathrm{c2,f}$ (red area) is again a region of pseudo-hysteresis,
where we see hysteresis in the numerics but hysteresis vanishes in the
presence of sufficient random fluctuations and in the experiment, 
while the yellow area is related to a genuine
hysteresis that should also be robustly observable
in an experiment in the presence of some fluctuations. 
This gives also rise to a shape hysteresis as indicated in
  Fig.\ \ref{fig:hysteresis_sketch}. The intermediate states
  upon snapping into the dome-like shape feature a flattened
  region around the center, while
  this feature is missing when the shape  continuously
flattens.

\section{Hydrodynamics}

\subsection{Model}

\begin{figure}[h]
	\centering
		\includegraphics[width=0.6\linewidth]{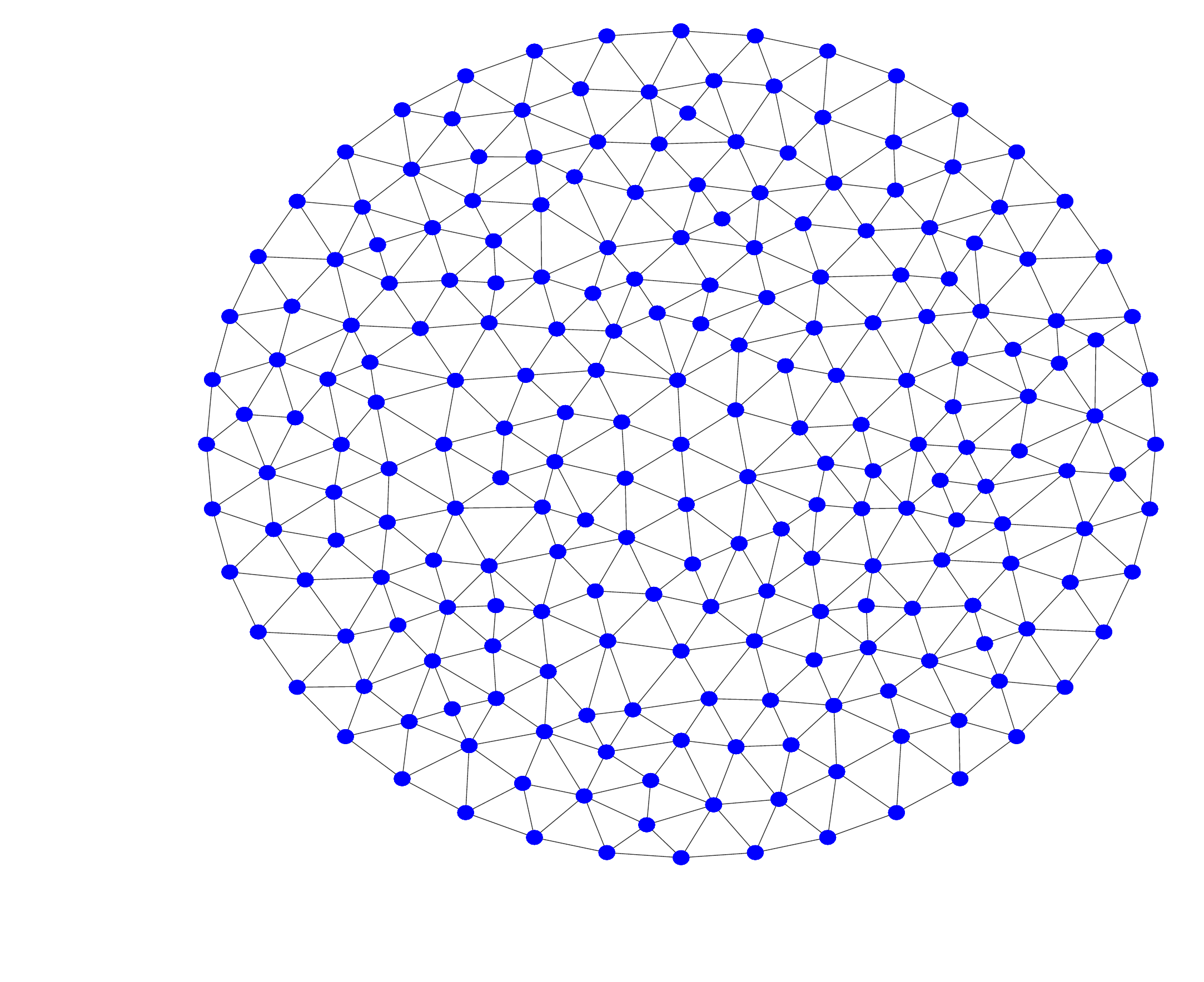}
        \caption{Illustration of the sphere mesh. Small spheres (blue
                  circles) are placed on the vertices of a Delaunay
                  triangulation. Example mesh with $n_\mathrm{B} = 40$
                  boundary vertices. The spheres have a radius
                  $a=2\pi R_\mathrm{out}/(10n_\mathrm{B})$, which is about
                  $10\,\%$ of their average distance.}
	\label{fig:sphere_mesh}
\end{figure}

In the following, we want to show that the hysteretic shape transition
of the modified framed elastic disk can be exploited 
as a propulsion mechanism for a microswimmer under a
periodic time-reversible driving of the swelling factor $\alpha$.
To this end, we need to
model the hydrodynamic interaction between the elastic disk and a surrounding
fluid. For this proof of concept we simulate the Stokesian
dynamics\cite{Durlosfsky1987}
and use the Rotne-Prager interaction\cite{Dhont1996,RotnePrager1969}. This
interaction describes the movement of a small sphere in the flow field of
another sphere. Therefore we  model our disk as a sheet of small
spheres and  place spheres of radius $a\ll R_\mathrm{out}$
on every vertex of the spring mesh, see Fig.\ \ref{fig:sphere_mesh}.  The
velocity $\vec{v}_i$ of every sphere $i$ can then be calculated from the
knowledge of the external forces $\vec{f}_j$ on all spheres $j$ via
\begin{align}
      \label{eq:v_rotne_prager}
      \begin{split}
        \vec{v_i} = \frac{1}{6\pi \eta a}\vec{f_i}+
        \sum\limits_{j\neq i} \frac{1}{6\pi \eta a}
        \left(\frac{3a}{4|\vec{r_i}-\vec{r_j}|}
          \left(\underline{\underline{\textbf{I}}} 
            + \frac{(\vec{r_i}-\vec{r_j})\otimes
         (\vec{r_i}-\vec{r_j})}{|\vec{r_i}-\vec{r_j}|^2}\right)\right.\\
   + \left.\frac{a^3}{4|\vec{r_i}-\vec{r_j}|^3}
     \left(\underline{\underline{\textbf{I}}}
       -3\frac{(\vec{r_i}-\vec{r_j})\otimes
         (\vec{r_i}-\vec{r_j})}{|\vec{r_i}-\vec{r_j}|^2}\right)
      \right)\vec{f_j}.
      \end{split}
\end{align}		
The constant $\eta$ describes the viscosity of the surrounding
fluid and $\underline{\underline{\textbf{I}}}$ represents the
three-dimensional unit matrix. In this model, we ignore torques
and rotations for simplicity.

The forces $\vec{f}_i$ can be calculated (analytically) from the
discretized stretching and bending energies \eqref{eq:stretch_energy_spring}
and \eqref{eq:EB_HK_discrete} as gradients with respect to
the vertex position $\vec{r_i}$
at each time step, and eq.\ \eqref{eq:v_rotne_prager} gives the resulting
vertex velocities.
The trajectory of each sphere and the disk's center
of mass as the average  of all sphere positions
are calculated by a simple Euler integration of the velocities with
a time step $\Delta t$.

\subsection{Simulation}

To simulate the movement of the disk, the trajectories of the spheres are
calculated based on the forces acting on them.
For a fixed swelling factor $\alpha$ this dynamics will relax
into the same force-free equilibrium state that we determined
also by  ``dry'' or static energy minimization in Fig.\
\ref{fig:energies_fixedFrame}. Using the dynamics  \eqref{eq:v_rotne_prager}
we can, however, obtain a realistic dynamics of each sphere position
and, thus, of the deformation and propulsion dynamics of the whole disk
in the presence of hydrodynamic interactions
in a viscous fluid. 

The general
concept of the simulation stays basically the same.
Again, $\alpha$ is changed
in small steps $\Delta \alpha$. After each step, the trajectories of the
spheres are calculated until the forces on the spheres fall below a threshold,
$\sum |\vec{f}_i| < \epsilon$. The simulation gives, in principle,
the corresponding
hydrodynamic time scale $\Delta \tau_\mathrm{h}$ on which this elastic
relaxation happens. 
That means that there are actually two
different time scales operating in this system. The swelling time scale
$T_\mathrm{sw}$ is defined by the swelling process and is  the time that
the disk needs to run through a complete deformation cycle.
This is the time scale  that can be externally controlled in an
experiment, where swelling frequencies $f_\mathrm{sw}
= 1/T_\mathrm{sw}\sim 5 {\rm Hz}$ are possible
for disks made from thermoresponsive hydrogels
if plasmonic heating of embedded gold particles by laser light
  is utilized
\cite{Mourran2017}. 
If we
divide a deformation cycle into $N$ small changes $\Delta \alpha_n$
and $\Delta \tau_\mathrm{h,n}$ is the 
hydrodynamic relaxation time for each step, we
obtain the second time scale
$\tau_\mathrm{h} = \sum_n \Delta
\tau_\mathrm{h,n}$, which is  the hydrodynamic relaxation time scale
for one deformation cycle and  determined by the interplay
of elastic forces and hydrodynamic friction. In
our simulation model we assume that hydrodynamic relaxation is much faster
than the swelling process,
$\tau_\mathrm{h} \ll T_\mathrm{sw}$, i.e.,
the disk swells slowly compared to its
deformation motion caused by the swelling, and we can use
quasi-equilibrated forces ($\sum |\vec{f}_i| < \epsilon$) along
the swelling cycle.

We can  estimate an order of magnitude for
the hydrodynamic relaxation time scale. The typical  
total force onto a  disk with 
shape height  $\Delta z$ close to  the instability is 
$F \sim \Delta z \sigma_0$ (see eq.\ \eqref{eq:stab}, which equates
areal force densities);
the disk has a friction coefficient $\sim \eta R_\mathrm{out}$, such
that the typical 
velocity  is
$\partial_t {\Delta z} \sim F/\eta R_\mathrm{out}\sim \Delta
z \sigma_0/\eta R_\mathrm{out}$, which 
leads to relaxation times
\begin{equation}
  \tau_\mathrm{h}
\sim\frac{ \eta  R_\mathrm{out}}{\sigma_0} \sim
\frac{\eta R_\mathrm{out}}{Y_\mathrm{3D}h(\alpha-1)} \sim
\frac{\eta \gamma_\mathrm{FvK}^{3/2}}{Y_\mathrm{3D}}\sim
    \frac{\eta R_\mathrm{out}^3}{Y_\mathrm{3D}h^3}
\label{eq:tauh}
\end{equation}
(using $\sigma_0 \sim
  Y_\mathrm{2D}(\alpha-1) \sim
       Y_\mathrm{2D}/\gamma_\mathrm{FvK}  $ close
  to the instability for an unframed disk, see eq.\ (\ref{eq:alphace})).
Typical elastic moduli  for
thermoresponsive hydrogels
are $Y_\mathrm{3D} \sim  10-100 {\rm kPa}$ \cite{Matzelle2003}
and F\"oppl-von K\'arm\'an numbers for the disks in Ref.\ \citenum{Mourran2017}
 are $\gamma_\mathrm{FvK} \approx 300$ 
(for $R_\mathrm{out} =30 {\rm \mu m}$ and $h = 5 {\rm \mu m}$)
and result in  fast hydrodynamic 
relaxation  time scales 
 $\tau_\mathrm{h} \sim   10^{-5} {\rm s}$ in water, such that
swelling frequencies up to $f_\mathrm{sw} \sim 10^5 {\rm Hz}$
still satisfy quasi force-equilibrium  along
the swelling cycle as assumed in our simulation.

On the other hand, 
the hydrodynamical relaxation time scale $\tau_\mathrm{h}$ should 
be large  enough (the hydrodynamically damped deformation or snapping
velocity of the disk slow enough)
for the
underlying  assumption of
low Reynolds number hydrodynamics to apply.
This is the case if
${\rm Re} \sim \rho  R_\mathrm{out}^2/\eta \tau_\mathrm{h}<1$ or 
$\tau_\mathrm{h} >  \rho R_\mathrm{out}^2/\eta $.
%
%
Inserting $\tau_\mathrm{h}$ from eq.\ (\ref{eq:tauh}) 
we obtain a condition  on the disk geometry, 
$ h^3/R_\mathrm{out} <  \eta^2/Y_\mathrm{3D}\rho \sim  10^{-1}{\rm \mu m}^2$,
where the last estimate is 
for the density and viscosity of water and moduli
 $Y_\mathrm{3D} \sim 10 {\rm k Pa}$.
This implies that disks have to be designed sufficiently
thin (and, thus, bendable) to remain at 
 low Reynolds numbers, which 
is  possibly a critical point for  experimental
realizations.
For disks of radius  $R_\mathrm{out} =30 {\rm \mu m}$, thicknesses
$h<2 {\rm \mu m}$ are required. 
As long as  the low Reynolds number
assumption applies, the swimming distance per deformation
cycle is independent of the time scale $T_\mathrm{sw}$ of the
swelling process and, thus, the deformation velocity. 
 The speed of shape changes affects the swimming velocity but 
does not affect the swimming distance
  as long as shape 
changes remain sufficiently slow that the low Reynolds number 
assumption holds.

The quality of our discretization and Stokesian dynamics
simulation scheme can be assessed by monitoring the
fluid flow
\begin{equation}
   j_S = \sum_i \int_{A_i}
   (\vec{u}(\vec{r}) - \vec{v}_i)\cdot \vec{n}_i
   \label{eq:jS}
\end{equation}
through the discretized surface (with unit normals $\vec{n}_i$),
which is given by the relative
velocity of the fluid flow $\vec{u}(\vec{r})$ with respect to the
disk vertex velocities $\vec{v}_i$. The fluid flow field
can be obtained from the Rotne-Prager interaction as 
\begin{align}
      \label{eq:u_rotne_prager}
      \begin{split}
        \vec{u}(\vec{r}) = 
        \sum\limits_{j} \frac{1}{6\pi \eta a}
        \left(\frac{3a}{4|\vec{r}-\vec{r_j}|}
          \left(\underline{\underline{\textbf{I}}} 
            + \frac{(\vec{r}-\vec{r_j})\otimes
         (\vec{r}-\vec{r_j})}{|\vec{r_i}-\vec{r_j}|^2}\right)\right.\\
   + \left.\frac{a^3}{4|\vec{r}-\vec{r_j}|^3}
     \left(\underline{\underline{\textbf{I}}}
       -3\frac{(\vec{r}-\vec{r_j})\otimes
         (\vec{r}-\vec{r_j})}{|\vec{r}-\vec{r_j}|^2}\right)
      \right)\vec{f_j},
      \end{split}
\end{align}		
while the velocity $\vec{v}_i$ of vertex $i$ is
given by (\ref{eq:v_rotne_prager})
and is essentially $\vec{u}(\vec{r}_i)$ with a regularization
of the $j=i$ term.
The quality of the approximation can be measured by the
dimensionless permeability $|j_S/j_P|$, where $j_P$ is 
the fluid stream through a theoretically perfectly permeable
surface that moves with a velocity $\vec{v}_i$  through
the fluid (i.e., setting $\vec{u}(\vec{r}))=0$ in eq.\ (\ref{eq:jS})).
Small permeabilities indicate that discretization and Stokesian dynamics
is a good approximation; ideally, we reach  $|j_S/j_P|=0$ because no fluid
should pass through the surface. 
For $a\approx 0.1 l_0$ (where $l_0=2\pi R_\mathrm{out}/n_\mathrm{B}$
is the typical rest length of a spring in the discretized mesh) we
find  surprisingly
small  permeabilities $|j_S/j_P| < 20\%$ for discretizations
with $n_B>100$ in view of the fact that less than $5\%$ of the surface
area are covered by spheres.

\subsection{Swimming motion of the snapping elastic disk}

\begin{figure}[t!]
		\centering
		\includegraphics[width=0.8\linewidth]{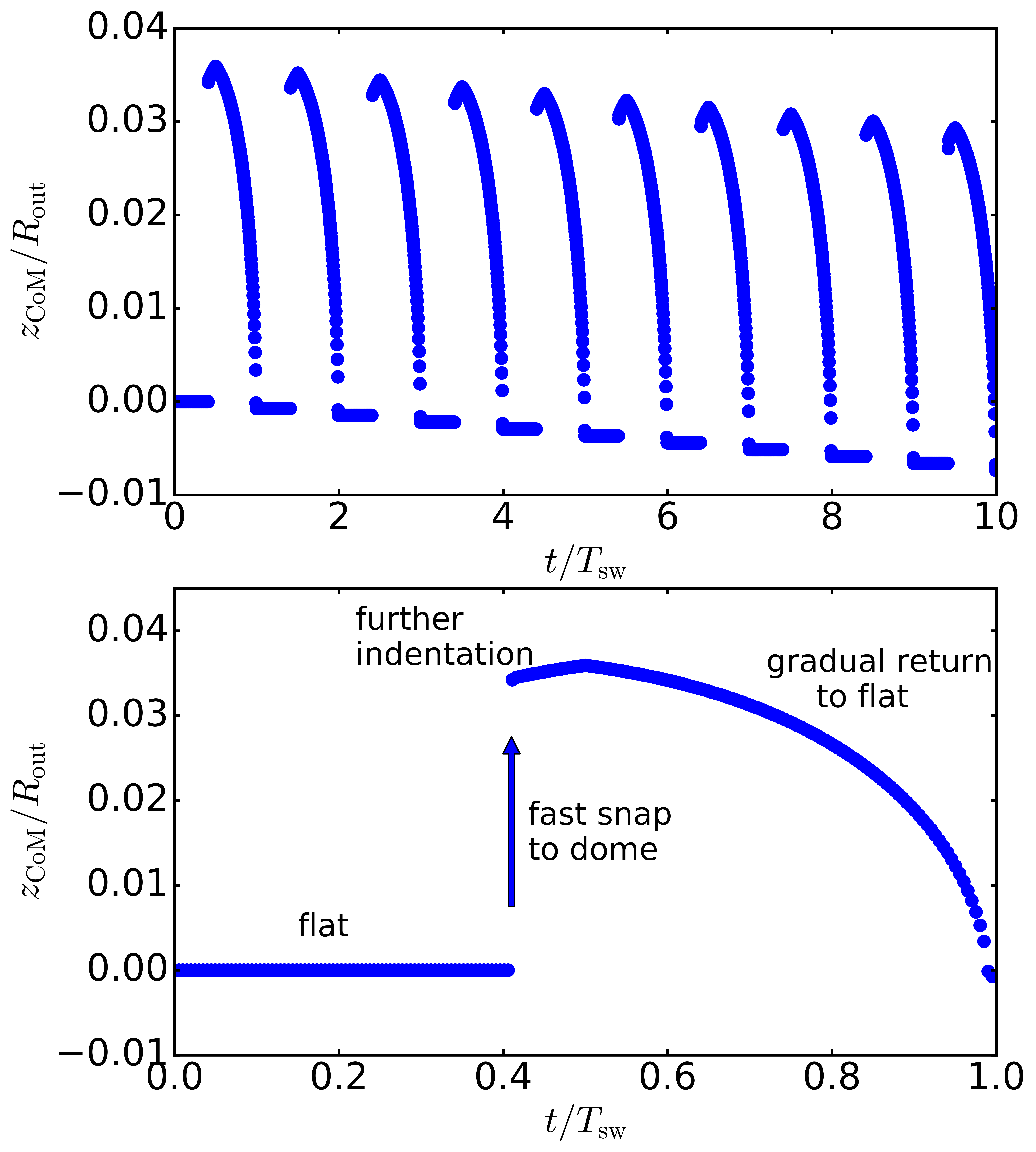}
		\caption{$z$-coordinate $z_\mathrm{CoM}$ of the disk's center
                  of mass as a function of time for ten full deformation
                  cycles. The simulated disk featured the same parameters as
                  the disk in Fig.\ \ref{fig:energies_fixedFrame} but with
                  $n_\mathrm{B}= 60$ and
                  $\varepsilon =
                  5\times 10^{-8}\,k_\mathrm{out}R_\mathrm{out}^2$. The spheres
                  in the mesh had a size of
                  $a=2\pi R_\mathrm{out}/(10n_\mathrm{B})$. In each 
                  swelling cycle, the swelling factor $\alpha$ varied between
                  $1$ and $1.1$ in 200 steps. The lower graph shows a zoom
                  into a single (the first) cycle.
                  The net swimming motion is in the direction of the
                opening.  }
		\label{fig:swimming_distance}
\end{figure}

In order to quantify the swimming motion of the disk
and proof the concept of a net swimming motion, we measure
the movement
of the disk's center of mass as a function of  time over multiple swelling
cycles.
Because of  its symmetry, the disk only moves into the direction
perpendicular to its initial plane, the $z$-direction. Therefore, Fig.\
\ref{fig:swimming_distance} shows the $z$-coordinate of the center of mass,
$z_\mathrm{CoM}$, as a function of  time for ten full swelling
cycles.
We will assume in the following
  that the dome-like shape always snaps downwards, i.e.,
  the opening is in negative $z$-direction as shown
  in Fig.\ \ref{fig:frame_sketch}.
In each
swelling cycle $\alpha(t)$  the swelling factor $\alpha$ changes  between
$\alpha_\mathrm{min}=1$ and $\alpha_\mathrm{max}=1.1> \alpha_\mathrm{c2,f}$
in 200 steps back and forth in a completely
time-reversible fashion.

Each swimming cycle consists of three phases
(see  Fig.\ \ref{fig:swimming_distance}).
Beginning at $\alpha=1$ and
$z_\mathrm{CoM} (t=0) = 0$, the disk does not move at
first as long as  the disk stays flat until $\alpha > \alpha_\mathrm{f,c2}$,
where the disk snaps into a dome-like shape in the numerics.
The second phase starts with 
the short snapping process itself.
During snapping  into the
dome-like shape, the swimmer moves into the positive $z$-direction
(the direction of the
tip of the elliptic dome, upwards in Fig.\ \ref{fig:frame_sketch}).
This movement happens nearly instantaneously on the swelling time scale.
Then, for $\alpha_\mathrm{f,c2}<\alpha<\alpha_\mathrm{max}$,
the height $\Delta z$
of the dome-shape slowly continues to grow. 
Then, the swelling factor starts to shrink again, and the third
phase starts. 
Here, the shape flattens again, and
we see a movement into the negative $z$-direction. When
the disk arrives back in its original state ($\alpha = 1$), a small net
motion into the negative $z$-direction, i.e.\ in the direction of the
                opening, remains,
which results in an effective
slope in the diagram over multiple swelling cycles in  Fig.\
  \ref{fig:swimming_distance}.

The effective 
speed of this swimmer is quite small
with a net swimming distance of just below
$1\,\%$ of the disks radius $R_\mathrm{out}$ after ten deformation cycles
for the parameter values given in Fig.\ \ref{fig:swimming_distance}.
This is a consequence of the known problem of all shape-changing
microswimmers that 
the swimming distance typically
scales only  quadratically with the deformation
displacement \cite{Lighthill1952,Blake1971}.
Nevertheless, this proves the principle
that a the hysteretic shape transition of a flat
elastic disk can be used as a propulsion mechanism for a microswimmer.
Shape-changing swimmers can still be effective if the
driving frequency (the swelling cycle frequency for our swimmer)
is sufficiently high. 
At low Reynolds numbers, the resulting swimming distance
$\Delta z_\mathrm{CoM}=z_\mathrm{CoM}(t=T_\mathrm{sw})$
is also independent of the shape of the
actual swelling cycle $\alpha(t)$  but can only depend on the values of 
$\alpha_\mathrm{min}=1$ and $\alpha_\mathrm{max}=\alpha_\mathrm{f,c2}$
characterizing the hysteretic part of the swelling cycle.

  We can use the  reciprocal theorem \cite{Stone1996} in order to obtain
  analytical insight into 
  the dependence of the  swimming  distance $\Delta z_\mathrm{CoM}$ on the
   deformation displacement $\Delta z$ of the swimmer.
We focus on the $z$-component as axisymmetric deformations can only
lead to motion along the axis of symmetry of the disk. 
For a deformed disk shape $A(t)$  with a 
shape deformation  $z=z(r,t)$ and a deformation velocity
$\dot{z}$ in $z$-direction ($r$ is the
radial coordinate in the $xy$-plane)
the reciprocal theorem gives
\begin{align}
  F_D  v_\mathrm{CoM} &= - \int_{A(t)} dA  (\vec{n}\cdot \sigma)_z
                              \dot{z}
\end{align}
where $F_D$ is the $z$-component of the
viscous drag force of  a disk with rigid shape $A(t)$
and velocity  $v_\mathrm{CoM}$ in $z$-direction,
$\sigma_D$ the corresponding stress tensor of the fluid,
and $ (\vec{n}\cdot \sigma)_z$  the
$z$-component of the normal stress of the fluid
onto the shape $A(t)$; these quantities
are related
via $F_D = \int_{A(t)}dA (\vec{n}\cdot \sigma_D)_z$.
This leads to 
\begin{align}
  v_\mathrm{CoM} &= - \frac{ \int_{A(t)} dA
                         (\vec{n}\cdot \sigma_D)_z \dot{z}}{F_D}
  \nonumber\\
  &= - 
    2\pi \int_0^{R_\mathrm{out}}dr r  [\sigma](r,t)
    (1+ (\partial_r z)^2)^{1/2} \dot{z}(r,t)
    \label{eq:vcom}
\end{align}
where
\begin{align}
  [\sigma](r,t) &\equiv    \frac{(\vec{n}\cdot \sigma)_z}{F_D}
         =  \frac{1}{2\pi R_\mathrm{out}}
    \frac{1}{(R_\mathrm{out}^2-r^2)^{1/2}}
    \left( 1- {O}\left( (\partial_rz)^2 \right)\right)
  \label{eq:sigma}
\end{align}
is the stress difference (divided by drag force) between
the two faces of a disk with rigid shape $A(t)$.
The first term in the last equation is the result for the 
flat disk (see Refs.\ \citenum{Gupta1957,Tanzosh1996,Felderhof2012}).
For weakly deformed rigid disks there are corrections; for the scaling of the
net swimming distance with the  shape height $\Delta z$ it is crucial
how these corrections scale with $\Delta z$.
For the total drag force 
$F_D = \int_{A(t)}dA (\vec{n}\cdot \sigma_D)_z$ 
the  corrections are quadratic, $-{O}\left( F_D(z/R_\mathrm{out})^2\right)$.
This can be shown by
interpreting the weakly deformed rigid disk as perturbed flat disk
and applying the reciprocal theorem \cite{Masoud2019}. Because
both the
fluid velocity field and the pressure only vary quadratically
in $z$  close to  the flat disk shape \cite{Tanzosh1996}, boundary
conditions to the flow and, thus, the
fluid stresses and the drag receive only quadratic corrections. 
This is also 
 supported by an exact result for the axisymmetric Stokes flow past
spherical caps of opening angle
$\beta$ and radius $R$ \cite{Dorrepaal1976}, which can also
be interpreted as weakly bent rigid disks.
The friction force $F_\mathrm{D,cap} =
\mu v_\mathrm{CoM} R(6\beta + 8\sin\beta + \sin(2\beta))$
reduces in the disk limit $\beta \ll 1$,  where 
$R_\mathrm{out} \approx R \beta $ and  $z/R_\mathrm{out}\approx
\beta/2$, to
$F_D \approx 16\mu  v_\mathrm{CoM}R_\mathrm{out}
(1-  (z/R_\mathrm{out})^2/6)$ in the first two leading orders. 
Therefore, we expect to find a leading order reduction  of the
stress jump (\ref{eq:sigma}), which is  quadratic  in $z/R_\mathrm{out}$
for a weakly deformed rigid disk. 
This  allows us to extract the scaling
of the swimming velocity as a function of the deformation function
$z(r,t)$ which describes the shape hysteresis.

\begin{figure}[t!]
		\centering
		\includegraphics[width=0.8\linewidth]{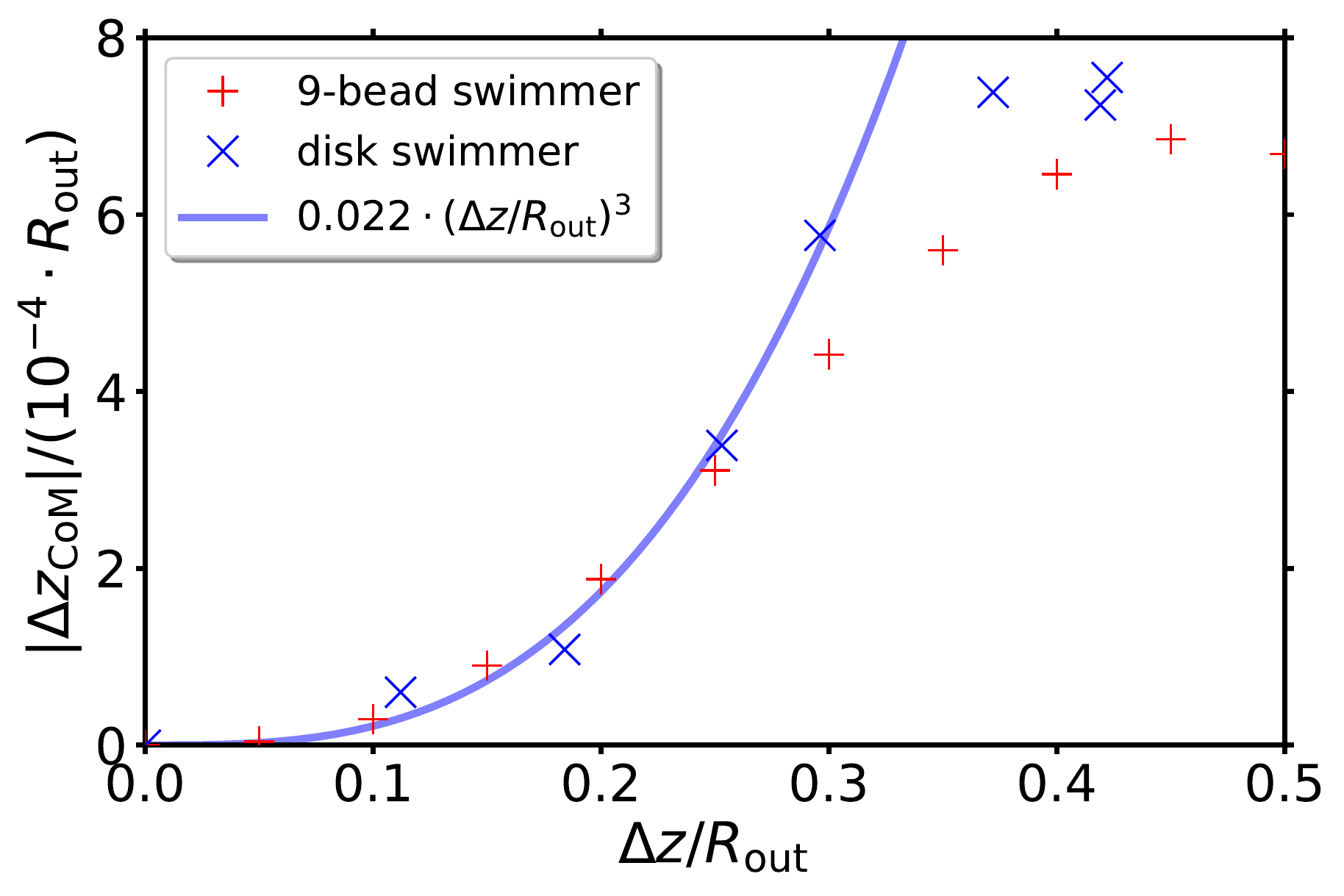}
		\caption{Net swimming distance  $|\Delta z_\mathrm{CoM}/R_\mathrm{out}|$ per
                  cycle  as a function of the shape height $\Delta z/R_\mathrm{out}$
                   both for  disks and 9-bead swimmer.
                  For the 9-bead swimmer the parameter $\Delta z$ is directly 
                  prescribed. 
                  For disks the increasing shape height  $\Delta z$ in the
                  snapped state
                  is realized by increasing the thickness
                  in the range $h/R_\mathrm{out}=0.02-0.14$, the other
                  parameters are as in  Fig.\ \ref{fig:energies_fixedFrame}.
                   The fit is $\Delta z_\mathrm{CoM}/R_\mathrm{out}
                   =  0.022(\Delta z/R_\mathrm{out})^3$.}
		\label{fig:zcom_deltaz}
\end{figure}

  The shape hysteresis as sketched in  Fig.\ \ref{fig:hysteresis_sketch}
  is mainly caused by flattened shapes during snapping ($\dot z(r)<0$,
  $v_\mathrm{CoM}>0$),
  while the shape remains dome-like during re-flattening  ($\dot z(r)>0$,
  $v_\mathrm{CoM}<0$).
   Expanding in the deformation $z$ 
   in eq. (\ref{eq:vcom}) and integrating over  time to obtain
   the swimming distance, 
   $\Delta z_\mathrm{CoM} = \int_0^{T_\mathrm{sw}}v_\mathrm{CoM}$,
   we realize that the leading order term
   only depends on final and initial state.
   It 
   gives the same  swimming distance
   $ \Delta z_\mathrm{CoM,snap}\sim -\Delta z$ both in the
   snapping and re-flattening phase and 
   cancels out exactly  for a full
   shape cycle. The next-to-leading term, however, gives a slightly
   bigger contribution in the re-flattening phase, which should scale
   as
   $\Delta z_\mathrm{CoM}/R_\mathrm{out}
   \sim  -(\Delta z/R_\mathrm{out})^3$.
 Therefore, 
 the net swimming distance is only a third 
 order effect in the hysteretic  shape height $\Delta z$.
 This is even smaller than the quadratic order observed, for example,
  for deforming spheres  \cite{Lighthill1952,Blake1971} and
  appears
 to be a hydrodynamic consequence of the disk geometry.
 This is confirmed in Fig.\ \ref{fig:zcom_deltaz}, where
 the net swimming distance for swimmers with different thicknesses 
    is shown as a function of their shape height in the snapped state.

     The shape height in the subcritical bifurcation with hysteresis
   is limited by the stretching factor $\alpha$  of the shape.
   Because a fiber along the diameter of the disk will stretch  by a factor
    $\alpha-1$, the change in height will scale as $\Delta
    z/R_\mathrm{out}  \sim (\alpha-1)^{1/2}$
    for a curved disk.
    For the hysteretic part of the deformation cycle, the relevant
      stretching factor is $\alpha = \alpha_\mathrm{c,f}$. 
    This results in a net swimming distance 
    \begin{equation}
      \frac{|\Delta  z_\mathrm{CoM}|}{R_\mathrm{out}} \sim
      \left(\frac{\Delta z}{R_\mathrm{out}}\right)^3
      \sim (\alpha_\mathrm{c,f}-1)^{3/2}.
      \label{eq:zcom}
      \end{equation}
   The swimming distance $\Delta z_\mathrm{CoM}$ per stroke
can be increased by increasing $\alpha_{c,f}-1$ and, thus, 
the width of the
yellow hysteresis area in Fig.\ \ref{fig:energies_fixedFrame}
which can be achieved, for example,  by increasing the thickness
 and, thus,
the bending modulus of the disk. This is explicitly
  demonstrated in Fig.\ \ref{fig:zcom_deltaz}.
The thickness dependence $\alpha_{c,f}-1\propto h^2$ results in $|\Delta
  z_\mathrm{CoM}|  \propto h^3$.

The deformation of the disk into a dome-like shape resembles
the deformation pattern of a scallop as envisioned by Purcell
\cite{Purcell1977};
our disk is, however, hysteretic, which enables swimming.
Interestingly, we find swimming into negative
$z$-direction, which is {\it  the direction away from the 
  tip of the dome}. This is the opposite swimming direction that
we expect from Purcells's high Reynolds number scallop, which moves
in the direction of its tip by  thrusting fluid out of the opening.
Our disk, on the contrary, will pull in fluid through the
opening because the size of the opening is fixed by the frame.
Also at high Reynolds numbers this leads to inertial
thrust in the direction of the opening.

\section{9-bead model}

\begin{figure*}[!t]
	\centering
		\includegraphics[width=0.7\linewidth]{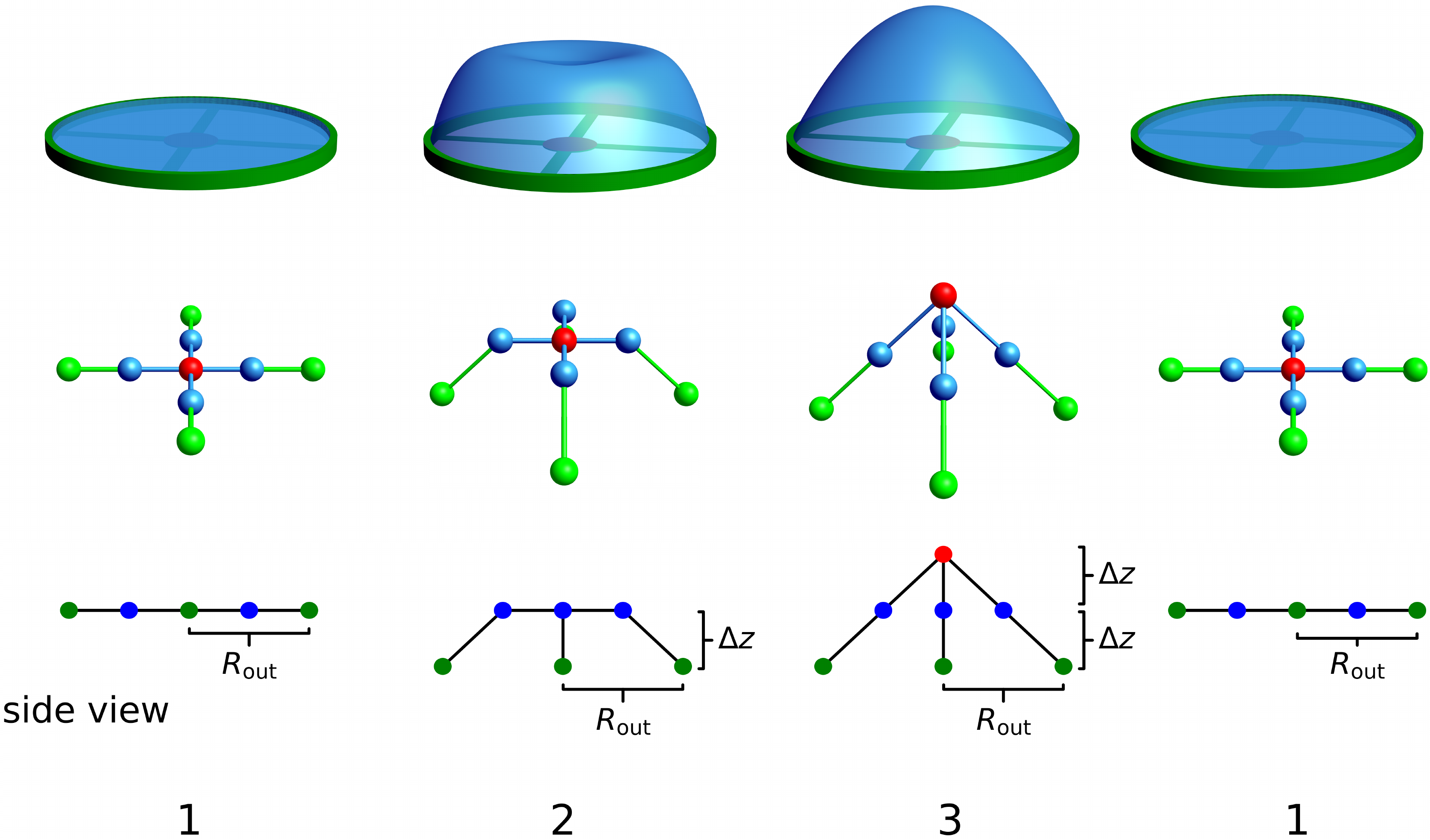}
                \caption{Illustration of the cyclic deformation
                  in the 9-bead model. First row:
                  corresponding states of the disk in the fixed frame. Second
                  row: three-dimensional sketch of the 9-bead swimmer
                  sates. Third row: two-dimensional side view on the 9-bead
                  model. Each column represents
                  one deformation state (1-2-3-1).}
	\label{fig:9-bead-illustration}
\end{figure*}

In order to further elucidate  the underlying propulsion mechanism, we
present a strongly simplified 9-bead model of the hysteretic scallop.
In this model, the swimmer only
consists of nine spheres that move on simple prescribed trajectories in
relative coordinates.
The 9-bead-model features three deformation phases, which mimic
the above three phases of the disk swimmer; these deformation phases are
visualized in Fig.\ \ref{fig:9-bead-illustration}. The sketches in the first
row show the corresponding states of the full, framed disk analogously to
Fig.\ \ref{fig:frame_sketch}. The second row shows three-dimensional
illustrations of the 9-bead swimmer and the last row a two-dimensional side
view.

The undeformed flat state is represented in the left column. Nine spheres are
aligned symmetrically in the $xy$-plane. The central sphere (red) is
surrounded by four spheres (blue) that are placed in a symmetrical way in a
distance of $R_\mathrm{out}/2$. The last four spheres (green) are set in
an identical way but in a distance of $R_\mathrm{out}$ to the central
sphere. They represent the fixed frame.  The connection lines between the
spheres indicate the structure of the object without a direct physical
meaning.  The second state (second column) represents a typical
transition state of
the disk during the snapping transition into the dome-like shape.
There, the outer regions
of the disk are already close to their final position, but the central region
does not show a tip yet. This is modelled by elevating the five inner spheres
(red and blue) in $z$-direction with $\Delta z$.
The third state then is the
finally stable elliptic dome-like shape after the transition. Here, the
central sphere is again elevated by another distance of $\Delta z$. Finally,
the last state is again the flat and relaxed disk so that the cycle can be
repeated. During the evolution from one state into another, the spheres move
on trajectories, which simply
linearly interpolate the positions of the spheres
between initial and final state. Therefore, the first evolution phase
($1\rightarrow2$) and the second phase ($2\rightarrow 3$) represent an
increasing swelling factor $\alpha$, while the last phase ($3\rightarrow 1$)
is related to a decreasing $\alpha$.

\subsection{Swimming behavior}

\begin{figure}[h!]
	\centering
	\includegraphics[width=0.8\linewidth]{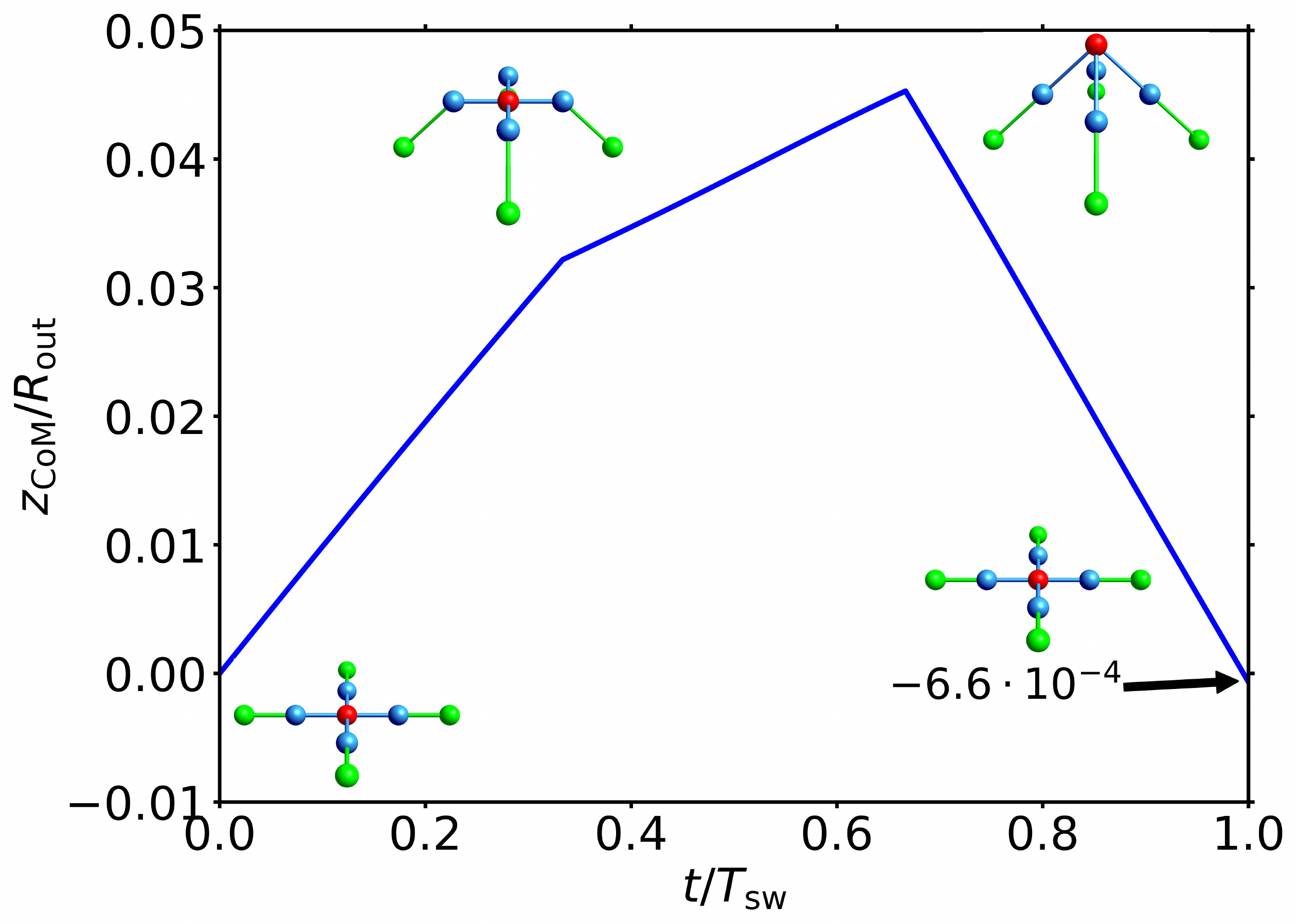}
	\caption{$z$-coordinate of the 9-bead swimmer's center of mass as a
          function of time during a single deformation cycle. The bead size is
          $a = 0.1\,R_\mathrm{out}$ and $\Delta z = 0.5\,R_\mathrm{out}$. The
          sketches illustrate the conformations of the swimmer during the
          deformation cycle.}
	\label{fig:9bead_swimming}
\end{figure}

The swimming motion of the 9-bead model is calculated with the same
hydrodynamic model as before. But now, the relative velocities of the spheres
are prescribed. So the forces on them have to be calculated. The knowledge of
eight relative velocities, together with the condition that the
swimmer is force-free, i.e., the total force
must vanish, enables us to calculate the motion of the center of mass
\cite{NajafiGolestanian2004,Golestanian2008}. The results are shown in Fig.\
\ref{fig:9bead_swimming} for a single deformation cycle.  In the first
deformation phase ($0<t< 1/3\,T_\mathrm{sw}$), the swimmer moves upwards in
positive $z$-direction. In the second phase, when the central sphere forms the
tip ($1/3\, T_\mathrm{sw}<t< 2/3\, T_\mathrm{sw}$), there is still an upwards
movement, which is simply slower than before because less spheres show a
relative motion in this phase. In the final phase ($t>\, 2/3 T_\mathrm{sw}$),
all spheres relax back to their original position causing a quick movement in
negative direction. After one complete cycle, at $t=T_\mathrm{sw}$, we see a
small net movement in negative direction with
$\Delta z_\mathrm{CoM} = -6.6\times 10^{-4}\,R_\mathrm{out}$. In
conclusion, we see the same qualitative behavior as for
the full disk model
in Fig.\ \ref{fig:swimming_distance}.
We can state that the simplified
9-bead model is able to model the swimming mechanism of a swelling disk that
has a fixed outer frame.
  For the 9-bead swimmer simulation results in Fig.\ \ref{fig:zcom_deltaz}
  show that the swimming  
 distance per cycle is also compatible with 
 $z_\mathrm{CoM}/R_\mathrm{out}\propto -(\Delta z/R_\mathrm{out})^3$
 for small  prescribed shape heights $\Delta z$ and collapses onto the
   results for the swelling disk.

\begin{table}[h]
\small
\caption{\ Swimmer parameter $\gamma$ (see eq.\ (\ref{eq:gamma})
  for 9-bead and disk
    swimmer in the beginning of the respective phases.}
  \label{tbl:gamma}
  \begin{tabular*}{0.48\textwidth}{@{\extracolsep{\fill}}lll}
    \hline
    phase   & $\gamma_\mathrm{9-bead}$  &  $\gamma_\mathrm{disk}$ \\
    \hline
    $1\to 2$ (neutral/weakly pushing,
         $\gamma$ small)  &  $-0.0068$   &  $0.39$\\
    $2\to 3$ (pulling, $\gamma<0$)  &  $-3.63$ &  $-10.88$ \\
    $3\to 1$ (pushing, $\gamma>0$) & $9.18$ & $1.24$ \\
    \hline
  \end{tabular*}
\end{table}

\begin{figure*}[p]
 \centering
      \includegraphics[width=1.0\linewidth]{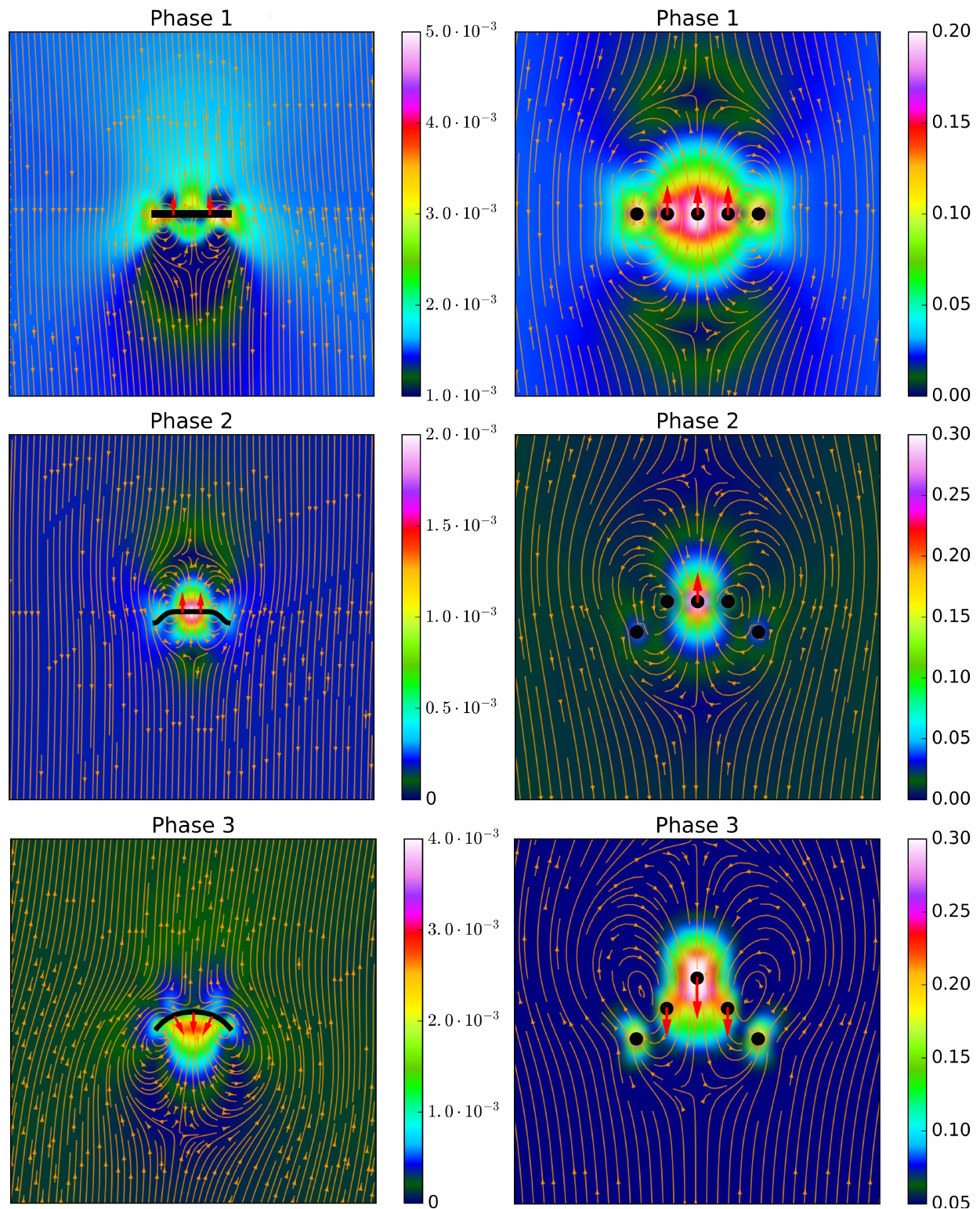}
      \caption{Velocity field $\vec{u}(\vec{r})$ in the $(y=0)$\,-plane for
        each deformation phase.  The left side shows the disk with a fixed
        frame from Fig.\ \ref{fig:energies_fixedFrame} and a discretization of
        $n_\mathrm{B}=40$. The right side shows a 9-bead swimmer with
        $a=0.1\,R_{\mathrm{out}}$ and $\Delta z =
        0.5\,R_{\mathrm{out}}$. Colors indicate the absolute $|\vec{u}|$ in
        units of $k_\mathrm{out}/\eta$ for the disk swimmer and in units of
        $R_\mathrm{out}/T_{\mathrm{sw}}$ for the 9-bead swimmer. The arrows
        show the direction of the fluid velocity vectors.}
      \label{fig:streamplots_combined}
    \end{figure*}

We can also compare the characteristics of the resulting fluid
flow field between the full disk model and the 9-bead model;
Fig.\ \ref{fig:streamplots_combined} shows that there
is good qualitative agreement between both models.
In particular, the characteristic stagnation point
  and the vortex ring  below the snapping disk  are
  reproduced. These features are characteristic for
  non-convex bodies and are also observed for dragged
  spherical caps \cite{Dorrepaal1976}.

We can expand the axisymmetric flow field  in the far-field
into Legendre polynomials \cite{Lighthill1952,Blake1971} and
extract the dipole contribution $p_2 P_2(\cos\theta)/r^2$ of the
radial part $u_r(r,\theta)$ of the flow field $\vec{u}(\vec{r})$
from eq.\ (\ref{eq:u_rotne_prager}) and normalize by the
center of mass velocity $|v_\mathrm{CoM}|$ to define
the dimensionless  parameter
\begin{equation}
  \gamma \equiv \frac{2p_2}{3R_\mathrm{out}^2|v_\mathrm{CoM}|}.
  \label{eq:gamma}
\end{equation}
Values $\gamma <0$ ($p_2<0$) indicate ``pulling'' motion of the
swimmer, values $\gamma>0$ ($p_2<0$)  ``pushing'' motion, and
$\gamma \approx 0$ ($p_2\approx 0$) a neutral motion \cite{Lauga2009}.
The results for $\gamma$ for both 9-bead and disk swimmer
are shown in  table \ref{tbl:gamma}.
In the beginning of phase $1\to 2$
(state 1 in Fig.\ \ref{fig:9-bead-illustration})
the disk is  approximately neutral, while the 9-bead swimmer
  is a weak pusher; the disk appears neutral because the
  subcritical
  snapping instability is caused by short-wavelength deformations.
In the beginning of
phase $2\to 3$
(state 2 in Fig.\ \ref{fig:9-bead-illustration}) disk and 9-bead swimmer
are pullers,
and in the beginning of phase $3\to 1$
(state 3 in Fig.\ \ref{fig:9-bead-illustration}) they are
pushers.

\section{Conclusions}

We pursued the concept of  a low Reynolds number swimmer 
based on the concept of utilizing  a hysteretic shape transition
in order to convert a
completely time-reversible oscillation of a
control parameter into a directed swimming motion.
We proved this  concept with 
 a flat circular elastic disk that  undergoes a 
 shape transition into  curved 
 shapes by a localized swelling
 of an inner disk or an exterior annulus. While
swelling the inner region of the disk with a constant swelling factor
$\alpha$ and the outer annulus with $1/\alpha$, we saw a transition into an
elliptic dome-like shape for $\alpha > 1$ and a transition into a hyperbolic
saddle shape for $\alpha < 1$.
The control parameter of this shape transition is the
swelling factor $\alpha$.
 We found the transition to be a supercritical bifurcation with only 
numerical hysteresis, which will  disappear in a real
experiment in the presence of some fluctuations.
We could re-establish a genuinely hysteretic shape transition by
replacing the outer annulus  by a fixed outer frame and by
introducing an additional
attractive short-range interaction in the central region.
The details of this interaction are not relevant, as long as it 
is an attractive short-range interaction, as it arises, for example,
from van der Waals forces  or screened electrostatic forces. 
We then see a hysteretic subcritical shape transition between a
flat state and a  dome-like  shape.

Embedding this framed disk into a viscous fluid at low Reynolds
numbers,  a Stokesian
dynamics simulation of the hydrodynamic interaction with
the surrounding fluid
with a time-reversible and cyclic changing of $\alpha$ showed
that the swimmer is effectively moving into
the direction of the opening of the dome.
This way, a self-deforming microswimmer can be realized that uses only a
single scalar
control parameter, the swelling factor $\alpha$. This control parameter
has to be changed only by a few percent in order to trigger a drastic
conformation change with the snapping transition into the curved shape.
Interestingly, the snapping into
an elliptic dome-like shape resembles  the opening
and closing of a
scallop as envisioned in the scallop theorem
\cite{Purcell1977}. As opposed to Purcell's scallop
the elastic  disk swimmer  actually performs directed motion at
low Reynolds numbers  because of the additional hysteresis.
 The swimming direction of  the snapping elastic disk  is 
  into the direction of the  opening of the dome (away from the tip).
  As for many  shape-changing low Reynolds number microswimmers
  swimming is a higher order effect in the deformation displacement. 
 For the snapping elastic disk 
  the net swimming distance per swelling cycle 
  is  only a third order effect in the
  height of the dome-like shape
  ($|\Delta z_\mathrm{CoM}/R_\mathrm{out}|  \sim
  -(\Delta z/R_\mathrm{out})^3$).
The swimming mechanism by hysteretic snapping is reproduced
  by a simplified 9-bead model of the disk. Qualitative agreement of the
  resulting flow fields, its pusher/puller characteristics and the
  net swimming distance (see figs.\ \ref{fig:zcom_deltaz} and
  \ref{fig:streamplots_combined}) show that the 9-bead model
    captures the essence of the swimming mechanism.

An experimental realization of the snapping disk
microswimmer concept could be possible 
using thermoresponsive hydrogels, which have already been
used for the implementation
of helical microswimmers \cite{Mourran2017,Zhang2017}.
These hydrogels can be swollen  by
  plasmonic heating of  embedded gold particles by laser light.
  Therefore, localized swelling is possible by embedding gold particles
  only in specific parts of the hydrogel, such that  these hydrogels
  are suited to realize the proposed disk microswimmers.
With hydrogel properties
from the literature, an estimate of the expected swimming speed is also
possible.  Mourran {\it et al.} investigated the thermal swelling behavior of
thermoresponsive hydrogels with the help of circular
disks\cite{Mourran2017}. These disks had a diameter of $30\,\mathrm{\mu m}$
and a thickness of $5\,\mathrm{\mu m}$. They showed cyclic diameter changes of
several percent caused by a pulsed laser with frequencies of
$1-5\,$Hz. Therefore, a frequency of $5\,$Hz seems to be realistic for the
realization of the deformation cycle of our microswimmer with comparable
dimensions. With a diameter of $30\,\mathrm{\mu m}$ and a swimming distance of
about $1\,\%$ of the radius after ten cycles, a net velocity of
$0.075\,\mathrm{\mu m}/s$ follows. 
Generally, shape-changing swimmers such as the snapping disk
can only  be made effective if the
driving frequency (the swelling cycle frequency for our swimmer)
is sufficiently high. On the one hand, we  expect higher
frequencies to be possible for thin disks.
On the other hand, 
the swimming distance per swelling cycle
 can be increased by increasing
the thickness of the disk ($|\Delta z_\mathrm{CoM}|  \propto h^3$).

One has to bear in mind that the snapping transition
dynamics has to be sufficiently slow for low Reynolds number
hydrodynamics to apply. We estimated the typical hydrodynamical
relaxation time scale, which is the typical time scale for snapping
in a viscous fluid, to be 
$\tau_\mathrm{h} \sim \eta R_\mathrm{out}/\sigma_0 \sim
 \gamma_\mathrm{FvK}^{3/2} \eta/Y_\mathrm{3D}$, which is 
 of the order of  $\tau_\mathrm{h} \sim 10^{-5} {\rm s}$   for
the thermoresponsive hydrogel disks from Ref.\ \citenum{Mourran2017}. 
Then ${\rm Re}<1$ is equivalent to 
$\tau_\mathrm{h} >  \rho R_\mathrm{out}^2/\eta $ and  restricts
the low Reynolds number regime to sufficiently thin  disks with 
$ h^3/R_\mathrm{out} <  10^{-1}{\rm \mu m}^2$.
At higher Reynolds numbers, we also
expect swimming motion in the direction away from the 
  tip of the dome.

The concept of a low Reynolds number swimmer 
based on the principle of  a hysteretic shape transition
can be applied to other snapping systems such as
tube-like  \cite{Overveldea2015} or
shell-based  snapping systems \cite{Gorissen2020}.
All systems with a  hysteretic snapping instability should give
rise to a net swimming motion 
in a viscous fluid at low Reynolds numbers.
As opposed to our disk swimmer, these tube and shell objects
enclose volume and employ 
 volume or pressure control, such that genuinely autonomous
swimming is not possible because swimming  requires attached
devices to control volume or pressure \cite{Djelloul2017}.
Swelling or shrinking materials as proposed in this work are
an alternative to control the equilibrium area and, consequently,
also equilibrium enclosed volume of these objects by swelling and shrinking,
which can often
trigger the same snapping transition at fixed actual volume.

\section*{Conflicts of interest}
There are no conflicts to declare.

\section*{Acknowledgements}

We acknowledge financial support by the Deutsche Forschungsgemeinschaft 
via SPP 1726 ``Microswimmers'' (KI 662/7-2).

\bibliography{literature}

\end{document}